\documentclass[twocolumn,aps,pra,amsmath,amssymb,superscriptaddress,bibnotes]{revtex4-2}

\usepackage{graphicx}
\usepackage{amsfonts}    
\usepackage{graphicx}   
\usepackage{verbatim}   
\usepackage{color}      
\usepackage{subfigure}  
\usepackage{hyperref}   
\usepackage{natbib}
\usepackage{booktabs}
\usepackage{threeparttable}
\usepackage{dcolumn}
\usepackage{multirow}	
\usepackage{subfigure} 
\usepackage{ulem}  
\usepackage{multirow, array, rotating}
\usepackage{bm}
\usepackage{bbm}

\newcolumntype{M}{>{$\vcenter\bgroup\hbox\bgroup}c<{\egroup\egroup$}}

\usepackage[T1]{fontenc}
\newcommand{\Rmnum}[1]{\uppercase\expandafter{\romannumeral #1}}  

\newcommand{\beq}{\begin{equation}}
	\newcommand{\eeq}{\end{equation}}
\newcommand{\bqa}{\begin{eqnarray}}
	\newcommand{\eqa}{\end{eqnarray}}

\newcommand{\rt}[1]{\sqrt{#1}\,}

\newcommand{\bra}[1]{ \langle{#1} |}
\newcommand{\ket}[1]{ |{#1} \rangle}

\newcommand{\sch}{Schr\"odinger}

\newcommand{\sq}[1]{\left[ {#1} \right]}

\newcommand{\tr}[1]{{\rm Tr}\sq{ {#1} }}

\newcommand{\id}{\mathbbm{1}}

\definecolor{maroon}{rgb}{0.7,0,0}

\definecolor{ngreen}{rgb}{0.3,0.7,0.3}

\definecolor{golden}{rgb}{0.8,0.6,0.1}

\begin{document}
	\title{Experimental verification of the steering ellipsoid zoo via two-qubit states}
	\author{Kai Xu}
	\affiliation{CAS Key Laboratory of Quantum Information, University of Science and Technology of China, Hefei 230026, People's Republic of China}
	\affiliation{CAS Center For Excellence in Quantum Information and Quantum Physics, University of Science and Technology of China, Hefei 230026, People's Republic of China}

	\author{Lijun Liu}
	\email{lljcelia@126.com}
	\affiliation{College of Mathematics and Computer Science, Shanxi Normal University, Taiyuan 030006, People's Republic of China}
	
	\author{Ning-Ning Wang}
	\author{Chao Zhang}\email{drzhang.chao@ustc.edu.cn}
	\author{Yun-Feng Huang}\email{hyf@ustc.edu.cn}
	\author{Bi-Heng Liu}
	\affiliation{CAS Key Laboratory of Quantum Information, University of Science and Technology of China, Hefei 230026, People's Republic of China}
	\affiliation{CAS Center For Excellence in Quantum Information and Quantum Physics, University of Science and Technology of China, Hefei 230026, People's Republic of China}
	\affiliation{Hefei National Laboratory, University of Science and Technology of China, Hefei 230088, People's Republic of China}
	
	\author{Shuming Cheng}
	\email{drshuming.cheng@gmail.com}
	\affiliation{The Department of Control Science and Engineering \& Institute for Advanced Study, Tongji University, Shanghai 201804, People's Republic of China}
	
	\author{Chuan-Feng Li}\email{cfli@ustc.edu.cn}
	\author{Guang-Can Guo}

	\affiliation{CAS Key Laboratory of Quantum Information, University of Science and Technology of China, Hefei 230026, People's Republic of China}
	\affiliation{CAS Center For Excellence in Quantum Information and Quantum Physics, University of Science and Technology of China, Hefei 230026, People's Republic of China}
	\affiliation{Hefei National Laboratory, University of Science and Technology of China, Hefei 230088, People's Republic of China}
	\date{\today}
	
	\begin{abstract}
		Quantum steering ellipsoid visualizes the set of all qubit states that can be steered by measuring on another correlated qubit in the Bloch picture. Together with local reduced states, it provides a faithful geometric characterization of the underlying two-qubit state so that almost all nonclassical state features can be reflected in its geometric properties. Consequently, the various types of quantum ellipsoids with different geometric properties form an ellipsoid zoo, which, in this work, is experimentally verified via measurements on many polarization-path photonic states. By generating two-qubit states with high fidelity, the corresponding ellipsoids are constructed to certify the presence of entanglement, one-way Einstein-Podolsky-Rosen steering, discord, and steering incompleteness. It is also experimentally verified that the steering ellipsoid can be reconstructed from using the twelve vertices of the icosahedron as measurement directions. Our results aid progress in applying the quantum steering ellipsoid to reveal nonclassical features of the multi-qubit system.
	\end{abstract}
	
	\maketitle
	

	\section{Introduction}
	
	If a bipartite quantum state is shared by two space-like separated parties, Alice and Bob say, then Bob's system can be steered to a specific state by Alice measuring on her part. This is the celebrated phenomenon of quantum steering that is first noticed by \sch~\cite{Schrodinger1935,Schrodinger1936} and subsequently generalized by others in several directions~\cite{Hughston1993,Gisin1996,Verstraete2002,Wiseman2007,Sania2014}. In particular, given a two-qubit state, Alice's steerability of Bob is fully captured by the quantum steering ellipsoid (QSE) which visualizes the set of all Bob's possible steered states in the Bloch picture~\cite{Sania2014}. Together with Alice’s and Bob’s local states, it provides a faithful geometric representation of the shared two-qubit state~\cite{Sania2014}, and thus generalizes the Bloch picture from the single qubit to two qubits. 
	
	The QSE reflects the rich structure of two-qubit states, which in turn induces a corresponding zoo of QSEs. Indeed, different geometric properties of the QSE have been explored to witness the hierarchical quantum correlations, such as Bell nonlocality~\cite{Milne2014B}, Einstein-Podolsky-Rosen (EPR) steering~\cite{Jevtic2015,Chau2016,Chau2016B,Quan2016,McClosky2017,Chau2017,Song2023}, entanglement~\cite{Sania2014,Chau2016,Milne2014}, and discord~\cite{Sania2014,Shi2011,Shi2012,Hu2015}. Moreover, it also appears to be useful for characterizing quantum coherence of steered states~\cite{Hu2016}, joint measurement reality~\cite{Hall2019}, quantum phase transitions~\cite{Du2021,Li2023,Rosario2023}, and monogamy of entanglement in new ways~\cite{Milne2014,Cheng2016, Zhang2019,Div2023,Song2023B}. 
	
	 Here, we experimentally verify the steering ellipsoid zoo via measurements on different photonic qubit states in the degrees of polarization and path. Specifically, QSEs are constructed to certify the presence of entanglement via the nested tetrahedral condition and genuine one-way EPR-steering respectively. The degenerated ellipsoids, including the pancake and needle, are also generated to witness discord. Furthermore, the subtle problem of whether the complete steering~\cite{Sania2014,Cheng2018} that all decompositions of Bob's reduced state can be steered to by Alice's one single measurement holds is examined, and the clear distinction between the steering completeness and incompleteness is drawn via the corresponding QSEs.
	 
	 It has been theoretically predicted in~\cite{Sania2014} and experimentally confirmed in~\cite{Zhang2019} that the set of all steered states for a two-qubit state forms an ellipsoid. Inspired by the fact that nine points are generically enough to determine an ellipsoid, we present an efficient approach to reconstruct the QSE by choosing the twelve vertices of the icosahedron as measurement directions. Our experimental results confirm that it is able to fit the theoretical-predicted QSE with a high precision.
	
	\section{Quantum steering ellipsoids}\label{QSE}
	
	Any two-qubit state $\rho_{AB}$, shared by Alice and Bob, can be written in the Pauli operator basis $\bm{\sigma}\equiv (\sigma_x, \sigma_y, \sigma_z)$ as
	\beq
	 \rho_{AB} = \frac{1}{4}(\id_A\otimes \id_B+ \bm{a} \cdot \bm{\sigma} \otimes \id_B+\id_A\otimes  \bm{b} \cdot \bm{\sigma} +  \sum_{i,j=x, y, z} T_{ij}\sigma _i\otimes \sigma_j). \label{state}
	 \eeq
	  Here $\id_A, \id_B$ denotes identity operators, $\bm{a}, \bm{b}$ the Bloch vectors of Alice's and Bob's local states, and $T\equiv(T_{ij})$ the spin correlation matrix. The measurement can be modeled as a positive-operator valued measure(POVM) $\{E_k\}$, with Hermitian operators $E_k$ satisfying $\sum_k E_k=\id$ and $E_k\geq0$ for all $k$. When a measurement is performed on Alice's qubit, each measurement outcome is associated to an element $E$ in a POVM, where there is $E  = e_0(\id+\bm{e}\cdot \bm{\sigma})$ with $0\leq e_0\leq 1$ and $|\bm{e}|\leq 1$ in the Pauli operator basis. Correspondingly, Bob's qubit is steered to 
	  \beq
	\rho_B^E=\frac{{\rm Tr}_A[\rho_{AB} E\otimes \id_B]}{p^E}= \frac{1}{2}\left[\id_B+\frac{(\bm{b}+T^\top\bm{e})\cdot \bm{\sigma}}{(1+\bm{a} \cdot  \bm{e})}\right] \label{steeredstate}
	 \eeq
	 with probability
	 \beq
	 p^E=\tr{\rho_{AB} E\otimes \id_B}=e_0(1+\bm{a} \cdot  \bm{e}). \label{probobability}
	 \eeq
	 
	 Considering all Alice's possible local measurements, it gives rise to a set of Bob’s steered states, represented by the set of Bloch vectors 
	\begin{equation}
		\mathcal{E}_{B|A} = \left\{\frac{\bm{b}+T^T \bm{e}}{1+\bm{a} \cdot \bm{e}}:~|\bm{e}|  \leq 1 \right\}. \label{qse}
	\end{equation}
	This set is proven to form an ellipsoid in the Bloch picture, called as quantum steering ellipsoid~\cite{Verstraete2002,Sania2014}. The subscript $B|A$ describes Bob’s steering ellipsoid generated by Alice’s local measurements, which is determined by its center
	\beq
	\bm{c}_{B|A}= \frac{\bm{b}-T^\top\bm{a}}{1-a^2}, \label{center}
	\eeq
	and its orientation matrix
	\beq
	Q_{B|A}=\frac{1}{1-a^2}(T-\bm{a}\bm{b}^\top)^\top(\id+\frac{\bm{a}\bm{a}^\top}{1-a^2})(T-\bm{a}\bm{b}^\top). \label{orientation}
	\eeq
	The eigenvalues and corresponding eigenvectors of $Q_{B|A}$ determine the squared lengths of the ellipsoid’s semiaxes and their orientations~\cite{Sania2014}. Here and elsewhere, we denote $x\equiv|\bm{x}|=\rt{\bm {x}^\top\cdot \bm{x}}$ for any vector $\bm{x}$. 
	
	Similarly, there is a steering ellipsoid $\mathcal{E}_{A|B}$ for Alice generated by Bob’s local measurements. It is worth noting that if the shared state is not symmetric under Alice and Bob, then Alice's ellipsoid is not identical to Bob's, which is confirmed in the following experiment on states $\rho_3, \rho_4, \rho_5$ and $\rho_8$ in Table~\ref{tab_states}.

	\section{The steering ellipsoid zoo}\label{SEZ}

	The set of Bob's steered state $\mathcal{E}_{B|A}$~(\ref{qse}) covers the whole Bloch ball if the state is a pure entangled state, i.e., $\ket{\psi_1}=\cos\theta\ket{00}+\sin\theta \ket{11}$ with $\theta \in (0, \pi/2)$ labelled as $\rho_1$ in Table~\ref{tab_states}, and reduces to a single point for $\theta=0, \pi/2$. The QSE can possibly vary from the $3$-dimensional ellipsoid (e.g. all entangled states) to $2$-dimensional ellipse (e.g. the separable state $\rho_7$ in Table~\ref{tab_states}), and to a straight line (e.g. the zero-discord state $\rho_6$ in Table~\ref{tab_states}). These suggest that its geometric properties have a close connection with the rich structure of two-qubit states.
	
	The above observations can be strengthened that together with Alice’s and Bob’s local Bloch vectors, the QSE yields a faithful characterization of the shared two-qubit state, up to local unitary operations~\cite{Sania2014}. Thus, it provides a powerful tool to reveal the nonclassical features underlying the state. For example, it gives a necessary and sufficient condition for the presence of entanglement via the nested tetrahedron condition that a two-qubit state is separable if and only if its steering ellipsoid fits inside a tetrahedron that itself fits the Bloch sphere~\cite{Sania2014}. This is verified on a family of Werner states~\cite{Werner1989} $\rho_2= p\ket{\psi_1}\bra{\psi_1}+(1-p) \id_A\otimes\id_B/4$ with $\theta=\pi/4$ and a varying $p\in [0, 1]$. It also yields a necessary and sufficient  condition for discord, measuring the difference of two natural quantum extensions of classical mutual information~\cite{Zurek2001,Henderson2001,Modi2012}, in the sense that Bob's ellipsoid becomes a segment of a diameter if and only if Bob has zero discord~\cite{Sania2014}, which is experimentally confirmed via the state $\rho_6$ in Table~\ref{tab_states}.

    The connections and distinctions among quantum steering, EPR-steering, and complete steering are also investigated in this work. Particularly, quantum steering refers to the phenomenon that Alice, by making suitable measurements, can steer Bob’s system to any desired state in the support of his local state~\cite{Schrodinger1936}, and her steerability of Bob is fully characterized by the steering ellipsoid as per~(\ref{qse}) for two-qubit states. EPR-steering describes the nonlocal phenomenon that one party can remotely prepare the other’s states with entanglement~\cite{Wiseman2007}, hence generalizing quantum steering as a kind of quantum correlations which lies strictly intermediate between Bell nonlocality and entanglement. Indeed, EPR-steerability can be fully determined by the corresponding QSE for a class of two-qubit states~\cite{Jevtic2015,Chau2016B,Quan2016}. Finally, the subtle problem of complete steering is about whether all decompositions $\{p_k, \rho_k\}$ of Bob's reduced state, i.e., $\rho_B=\sum_k p_k\rho_k$, can always be steered to by Alice's one single measurement such that $\rho_k=\rho^{E_k}_B$ and $p_k=p^{E_k}$  satisfying Eqs.~(\ref{steeredstate}) and~(\ref{probobability}) for each outcome $k$ and a given state $\rho_{AB}$~\cite{Sania2014,Cheng2018}. Its characterization and quantification have been thoroughly studied in~\cite{Cheng2018} via the QSE. Interestingly, all these steering properties are not symmetric under party permutations, and, as listed in Table~\ref{tab_states}, the state $\rho_3$~\cite{Bowles2016} is able to witness one-way EPR-steering and the state $\rho_5$~\cite{Sania2014} for one-way complete steering, both of which also automatically implies the asymmetry of quantum steering. More details are given in Supplementary Material~\cite{SM}. 

		\begin{figure*}[htbp]
		\centering
		\includegraphics[scale=0.22]{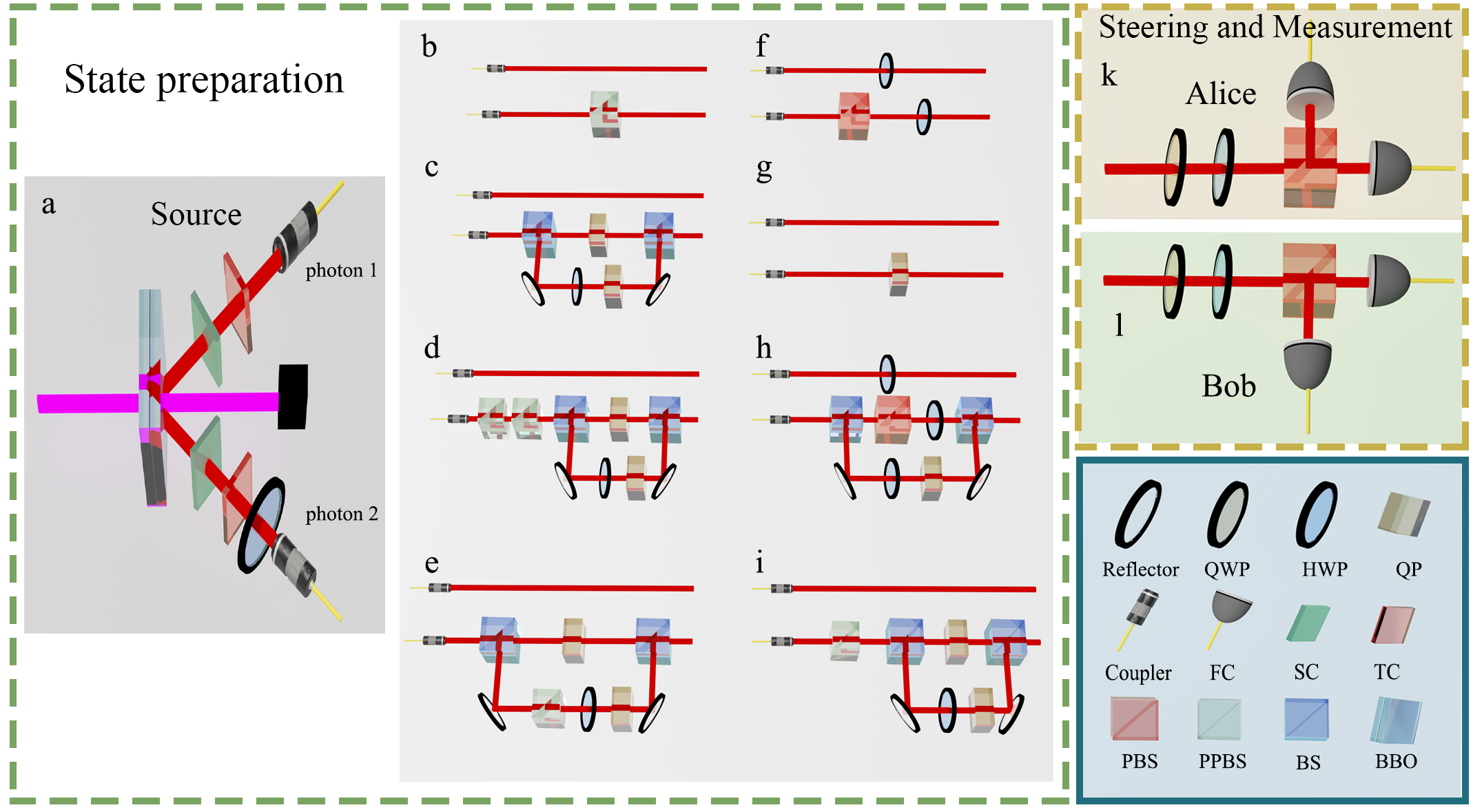}
		\caption{{\bf Experimental setup}. (\textbf{a}) The entanglement source. An ultraviolet laser pulse (centered at 390 nm) is used to pump a sandwiched beamlike type \Rmnum{2}  EPR sources, and a true-zero-order half wave-plate (THWP) is inserted between two $2$mm-BBOs that have the same optic axes. The spatial (LiNbO$_3$) and temporal (YVO$_4$) compensation crystals are employed to make this pair of photons indistinguishable. One photon is sent through a half wave plate rotated at $45^{\circ}$ to prepare the maximally entangled state $(\left | 00 \right \rangle +\left | 11 \right \rangle )/\sqrt[]{2} $. (\textbf{b})-(\textbf{i}) The preparation process of eight states in Table~\ref{tab_states}. Since \textbf{c}, \textbf{d}, \textbf{e}, \textbf{h} and \textbf{i} require mixing two states with a certain probability, two attenuation plates (which not drawn in the picture) are inserted into the two light paths which are further separated by the BS to adjust the relative light intensity. (\textbf{k})-(\textbf{l}) The steering and measurement process. These two operations admit the same ligh-path configuration which are composed of a rotatable half-wave plate, a rotatable quarter-wave plate, a PBS, and two fiber-coupled detectors. All rotatable wave plates are mounted on a motorized rotation stage. Symbols used in the figure are: HWP, half-wave plate; QWP, quarter-wave plate; QP, quartz plate;  FC, fiber-coupled detector; FC, spatial compensation crystal; TC, temporal compensation crystal; PBS, polarization beam splitter; PPBS,  partially polarization beam splitters; BS, beam splitter; BBO, barium borate crystals. }\label{setup}
	\end{figure*}

		\section{Experimental setup}\label{experimentalsetup}
		
		 To experimentally verify the zoo of quantum steering ellipsoids and reveal their nonclassical state features discussed above, we prepare a series of two-qubit states which are summarized in Table~\ref{tab_states}. In addition to the states explicitly mentioned in previous Sections~\ref{QSE} and~\ref{SEZ}, we also generate the state $\rho_7$ for the steering pancake and $\rho_8$~\cite{Cheng2018} for steering incompleteness in the ellipse case.
		
	   	\begin{table}[htbp]
	   	\caption{\label{tab_states} A series of two-qubit states used to verify the steering ellipsoid zoo. $\ket{\psi_-}=(\ket{01}-\ket{10})/\rt{2}$ in the Werner states $\rho_2$, $\ket{\psi_+}=(\ket{01}+\ket{10})/\rt{2}$ in $\rho_4$, $\rho_\theta={\rm Tr}_B[\rho_1]$ in $\rho_3$, and $\ket{+}=(\ket{0}+\ket{1})/\rt{2}$ in states $\rho_5$ and $\rho_7$.  }
	   	\begin{ruledtabular}
	   		\begin{tabular}{cl}
	   			& $\rho_1 =|\psi_1\rangle \langle\psi_1|,~\ket{\psi_1}=\cos\theta\ket{00}+\sin\theta \ket{11}$, $\cos\theta=\rt{2/3}$.   \\
	   			& $\rho_2 = p\left |\psi_-\right \rangle \left\langle\psi_-\right| +(1-p)\id_A\otimes\id_B/4, p=1/2, 1/3, 1/5.$                 \\
	   			&  $\rho_3 = p \ket{\psi_1}\bra{\psi_1} +(1-p)\rho_{\theta}\otimes \id_B/2, \theta = 0.3, p=0.55.$                                                                                          \\
	   			&  $\rho_4 = (\ket{00} \bra{00} +\ket{11}\bra{11}+2\ket{01}\bra{01}+4\ket{\psi_+}\bra{\psi_+})/8. $                                                                           							\\
	   			&  $\rho_5 = (\left | 00 \right \rangle \left\langle00\right| +\left | +1 \right \rangle \left\langle +1 \right| ) /2.$  \\
	   			&   $\rho_6 = (\left | 00 \right \rangle \left\langle00\right| +\left | 11 \right \rangle \left\langle 11 \right| )/2. $ \\                   
	   			&       $\rho_7 =(  \left | 0 0 \right \rangle \left\langle 00 \right| +  \left | 11  \right \rangle \left \langle 11 \right|+ \left | + + \right \rangle \left\langle ++ \right|)/3. $													\\
	   			&    $\rho_8 =(3 \id_A\otimes\id_B+\sigma _z\otimes \id_B+\sigma _x\otimes \sigma _x +\sigma _y\otimes \sigma _y)/12. $  \end{tabular}                                                                                              
	   	\end{ruledtabular}
	   \end{table}

		The experimental setup to generate these states in Table~\ref{tab_states} is displayed in Fig.~\ref{setup}. First, a maximally entangled state $(\left | 00 \right \rangle +\left | 11 \right \rangle )/\sqrt[]{2} $ is generated through the spontaneous parametric down-conversion (SPDC) process by pumping a sandwiched beamlike type-\Rmnum{2} Entanglement source~\cite{Zhang2015}. The fiber couplers are then used to transmit the pair of entangled photons into the state preparation optical path, where the photon going through the up path is labelled as photon $1$ and the down path as photon $2$ in (\textbf{a}). The light gray part of Fig.~\ref{setup} (\textbf{b})-(\textbf{i}) describe the preparation process of eight states listed in Table~\ref{tab_states} (see Supplementary Material~\cite{SM}  for more details about these states and their experimental preparation). Finally, the steering and measurement process is shown in the yellow dashed box of Fig.~\ref{setup}. Alice (Bob) randomly picks up a point on the Bloch sphere as the measurement direction, and then Bob (Alice) does tomography on his (her) steered state. After Alice (Bob) sampling all possible points for steering, all Bob's (Alice's) steered states are predicted to form an ellipsoid $\mathcal{E}_{B|A}$ as per Eq.~(\ref{qse}) ($\mathcal{E}_{A|B}$).

		\begin{figure*}[htbp]
			\begin{center}
				\begin{ruledtabular}
				\begin{tabular}{MMMMMM}
					\centering
					   &$\rho_1$&$\rho_{2, p_1}$&$\rho_{2,p_2}$& $\rho_{2, p_3} $&$\rho_3$\\
					\hline
					Type			&Ent. \& Comp. &Ent. \& Comp.    &	Sep. \& Comp.    & Sep. \& Comp.    &Ent. \& Comp   \\
			$\mathcal{E}_{A|B}$	 &\includegraphics[width=1in]{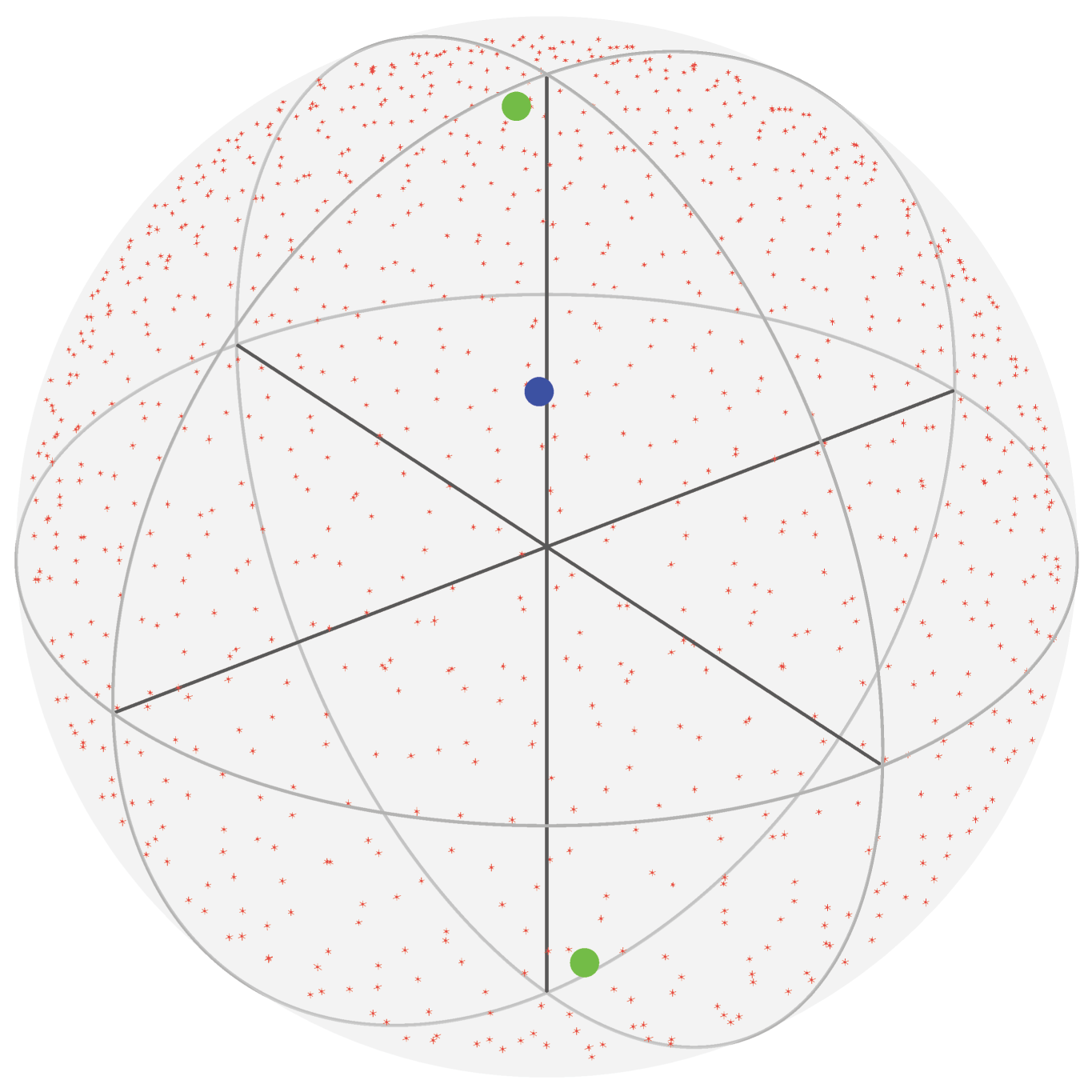}	&\includegraphics[width=1in]{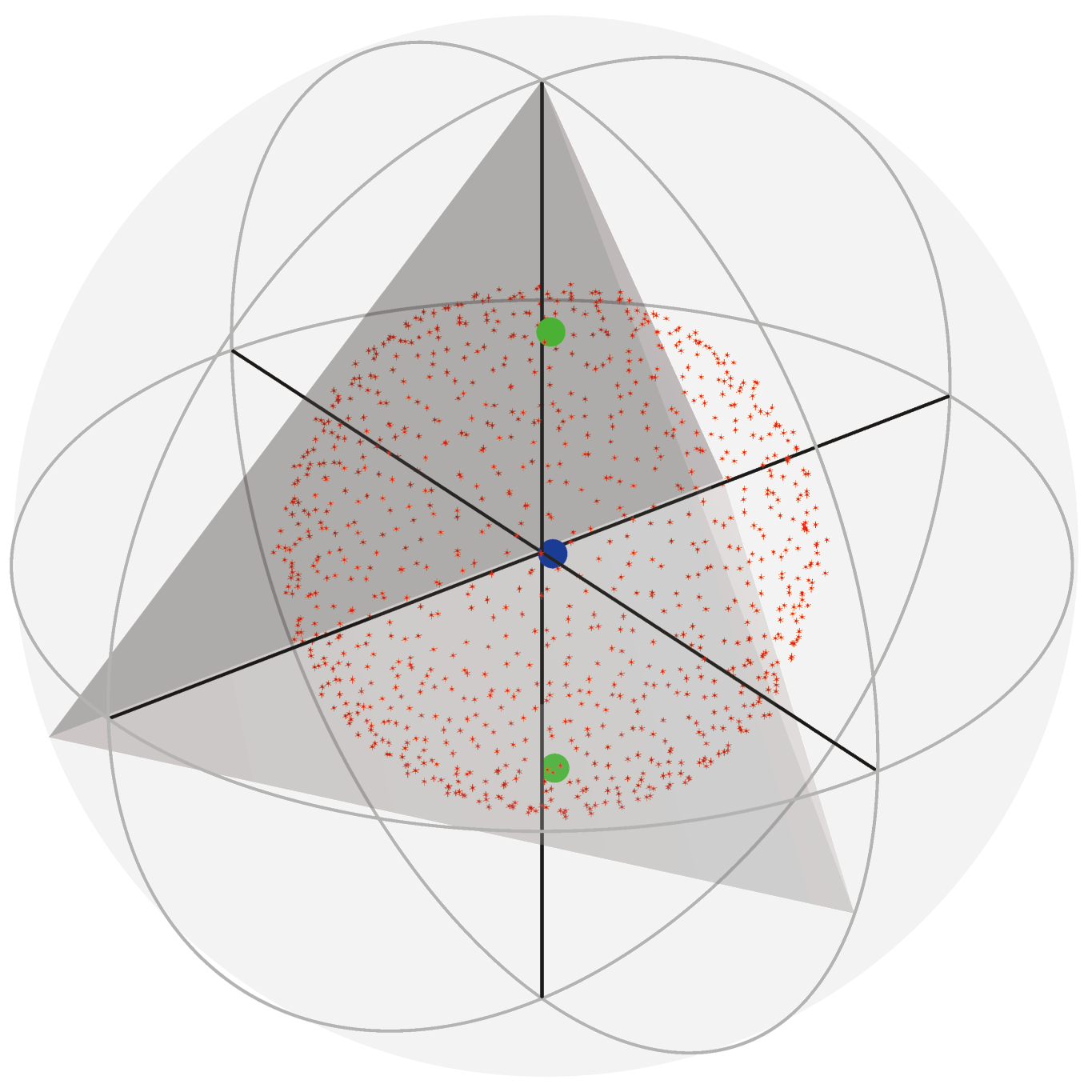}	&\includegraphics[width=1in]{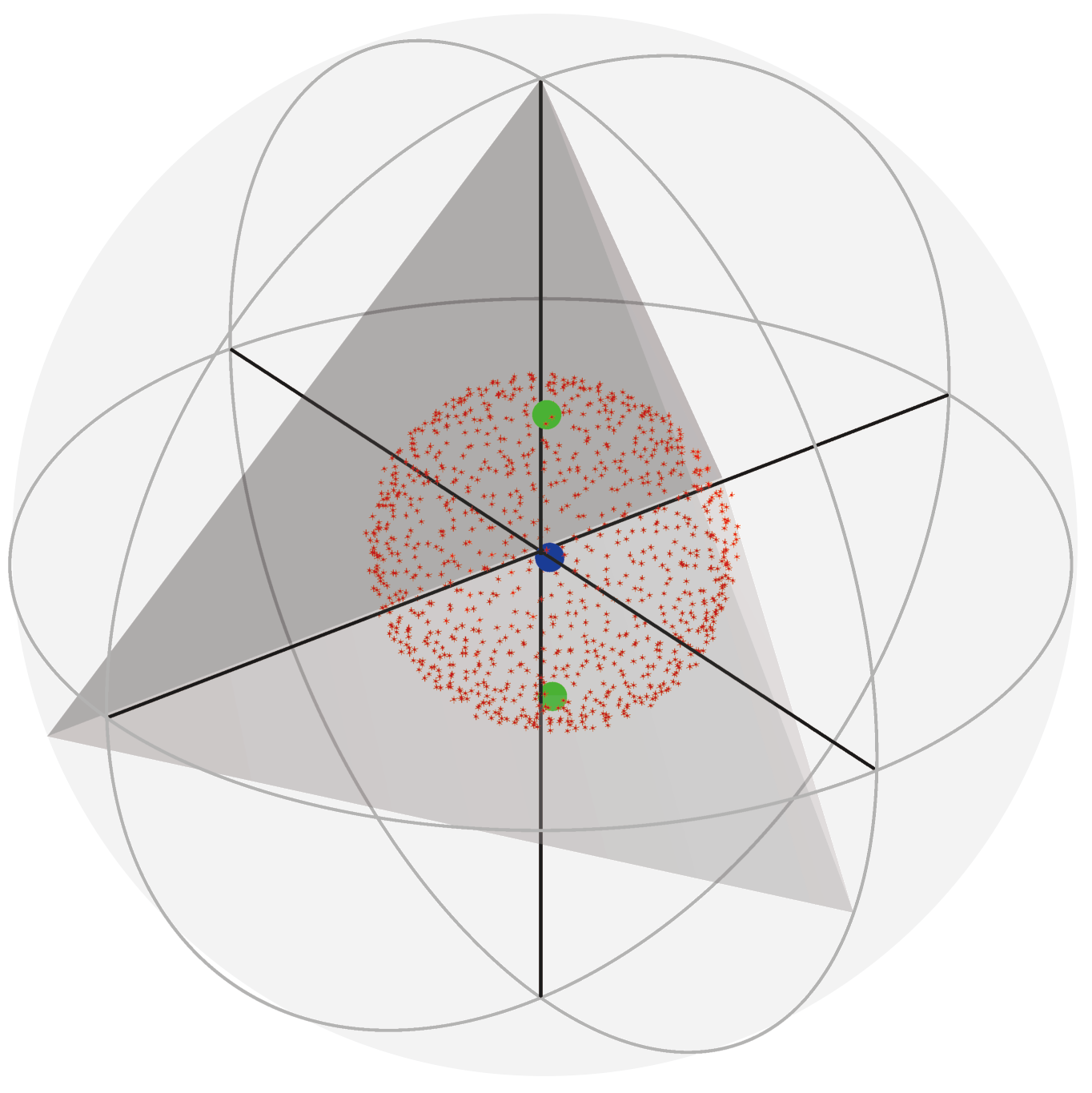}	&	\includegraphics[width=1in]{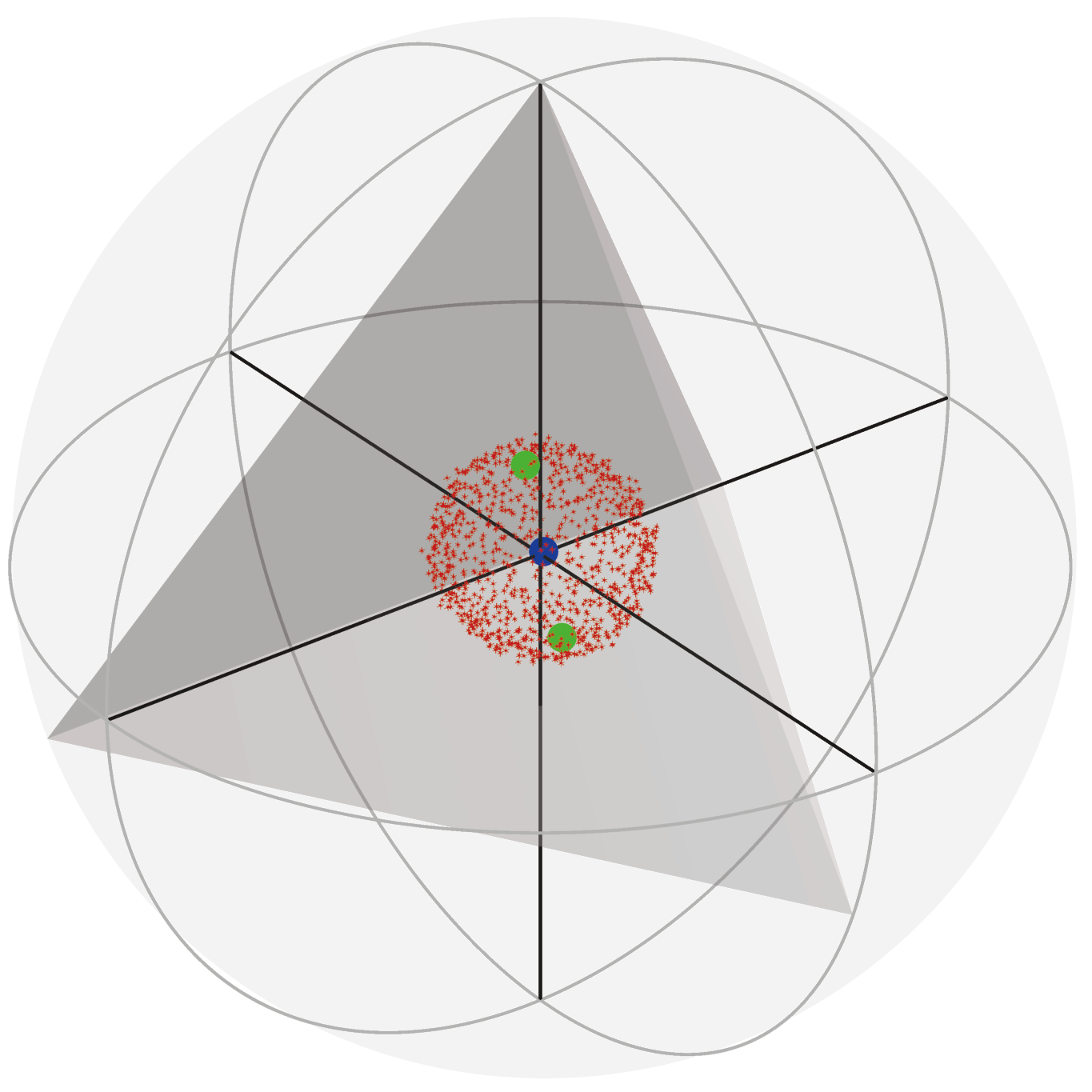}&\includegraphics[width=1in]{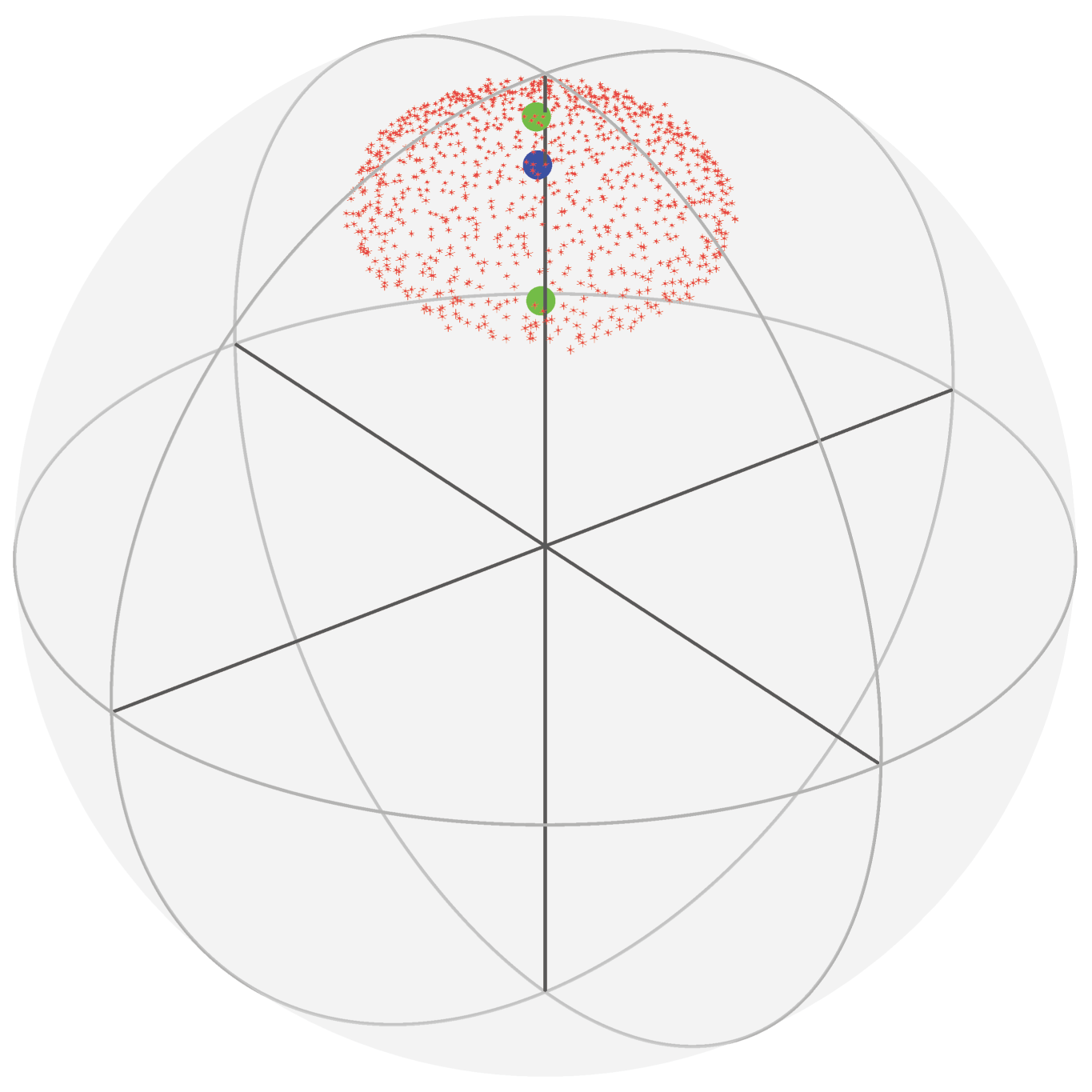}\\
				$\mathcal{E}_{B|A}$	& \includegraphics[width=1in]{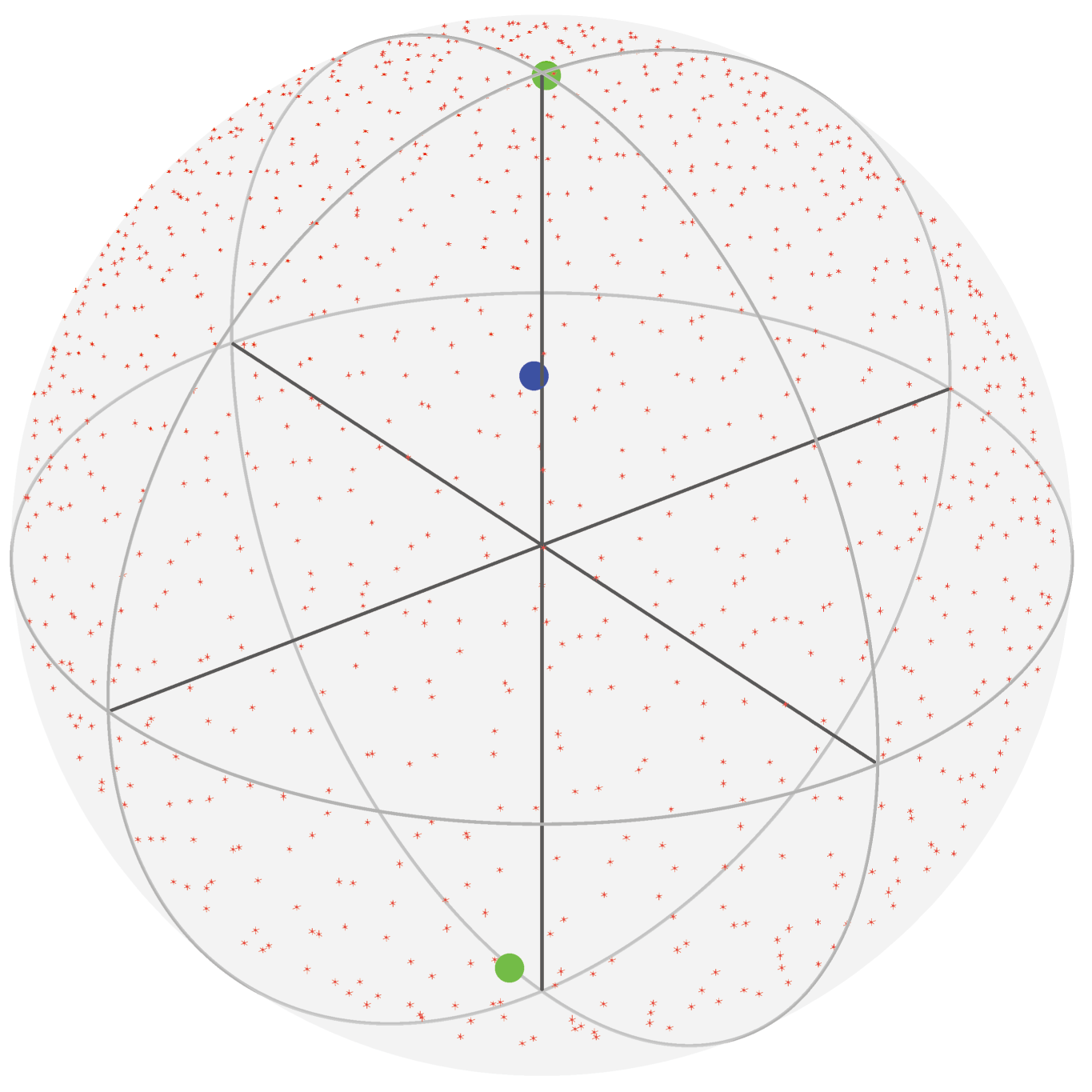} &\includegraphics[width=1in]{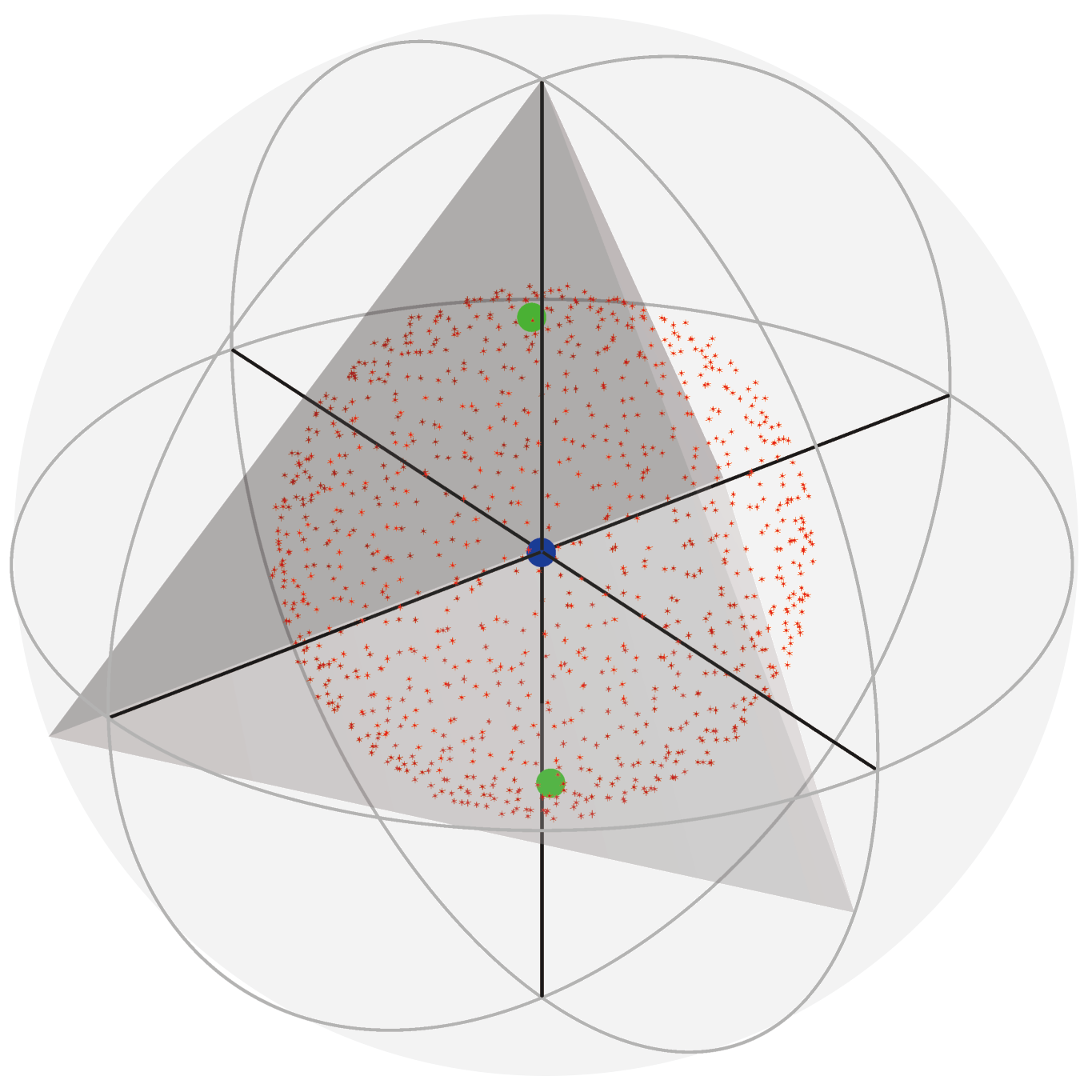} &  \includegraphics[width=1in]{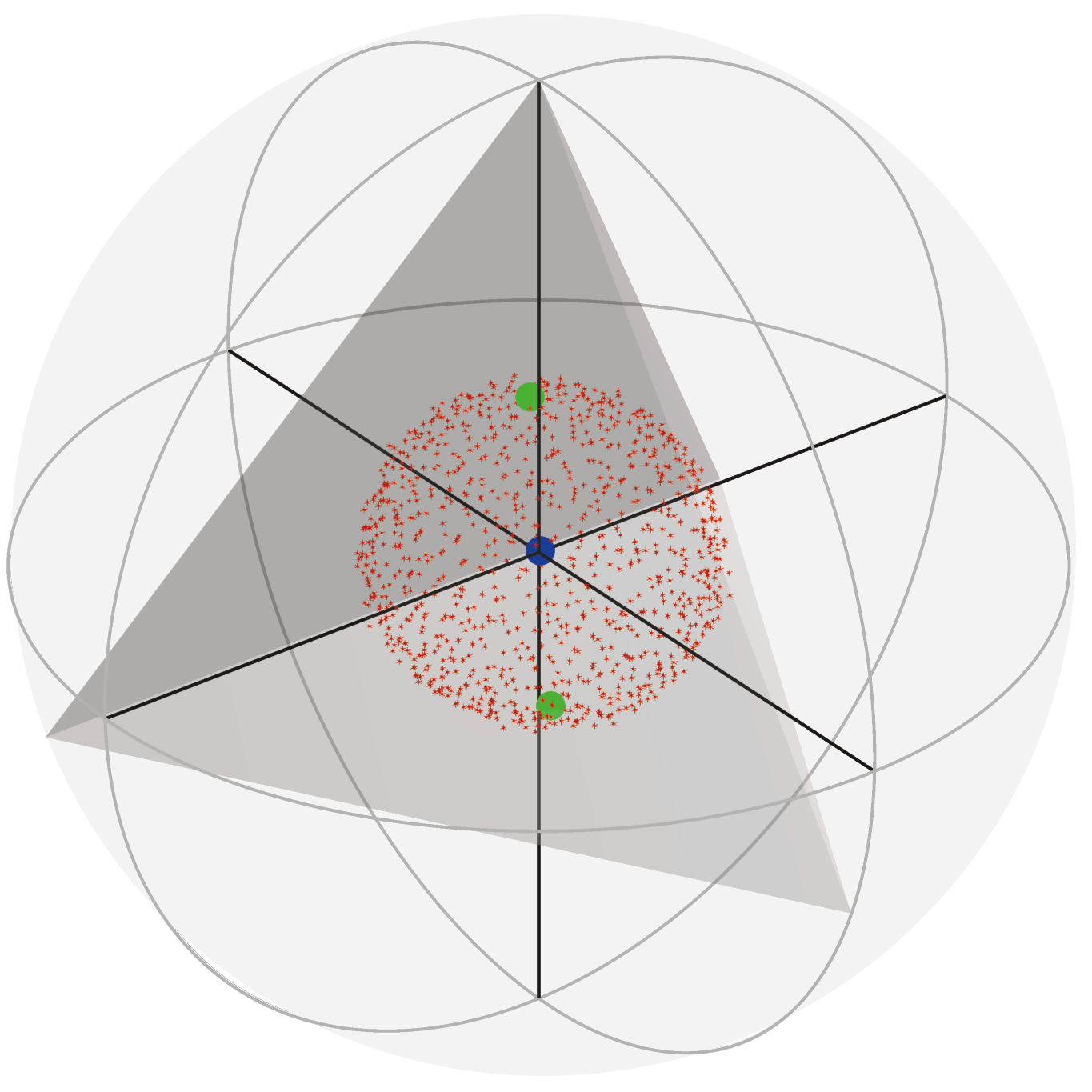}	&\includegraphics[width=1in]{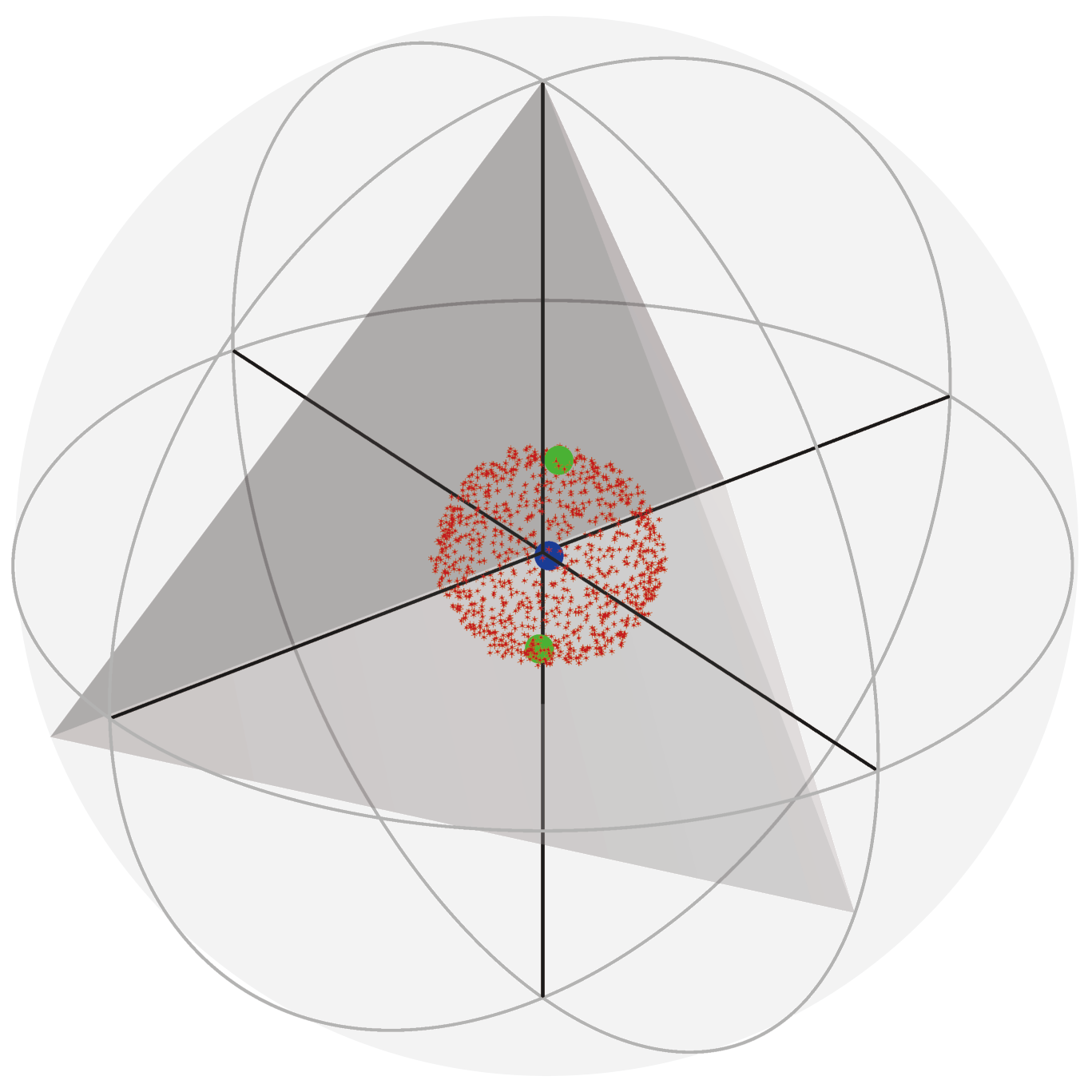}&\includegraphics[width=1in]{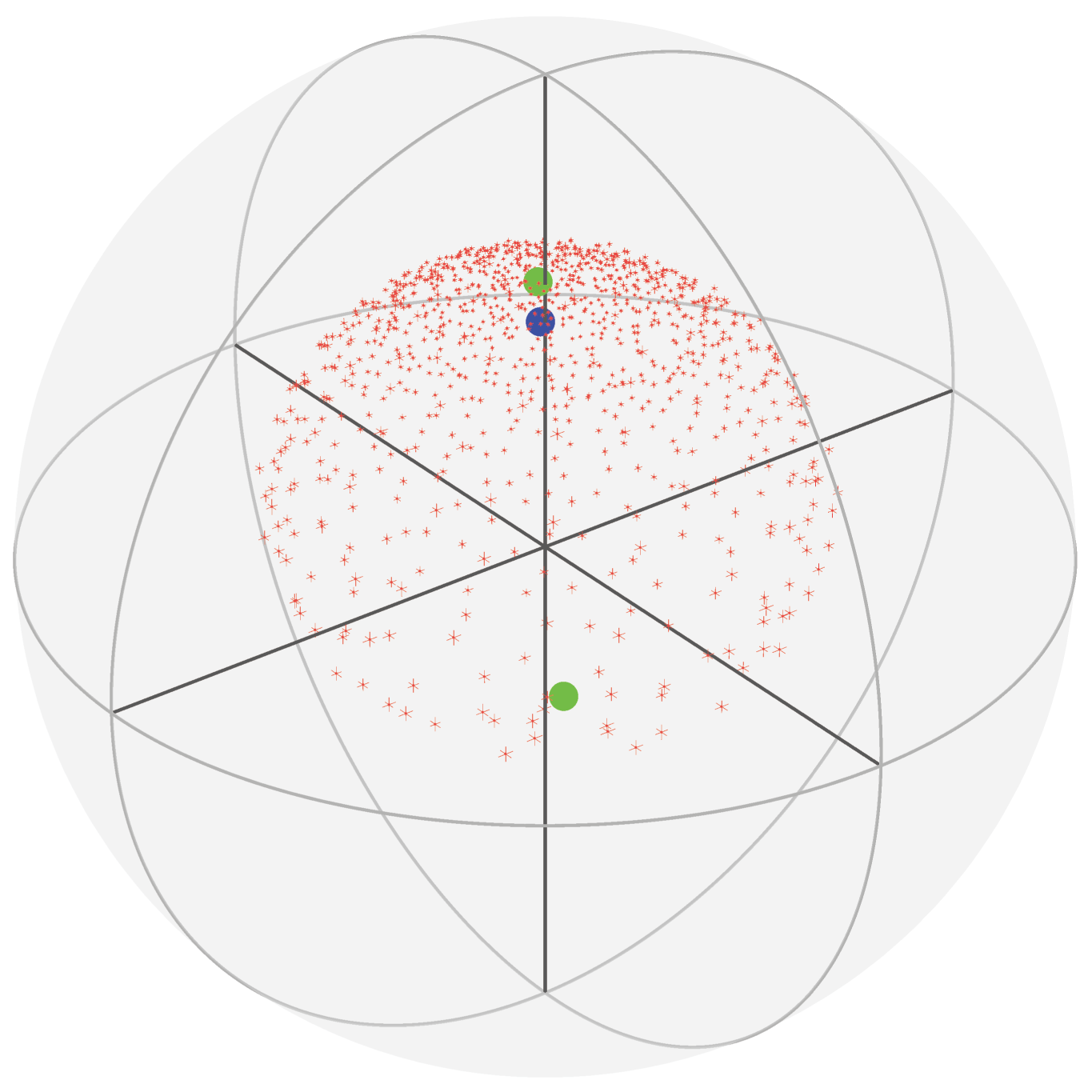}\\
				Con. & 0.8786$\pm0.0013$ (0.9428) & 0.2512$\pm 0.0015 $(0.2500)&0.0040$\pm 0.0015$(0)  &0(0)&0.1437$\pm 0.0018$(0.1837) \\
				\end{tabular}
			\begin{tabular}{MMMMMM}
				&$\rho_4$&$\rho_5$&$\rho_6$& $\rho_7 $&$\rho_8$\\
				\hline
				Type			&Ent. \& Comp.&Sep. \& Incomp.  &	Sep. \& Comp. &Sep. \& Comp.   & Sep. \& Incomp.  \\
			$\mathcal{E}_{A|B}$ &\includegraphics[width=1in]{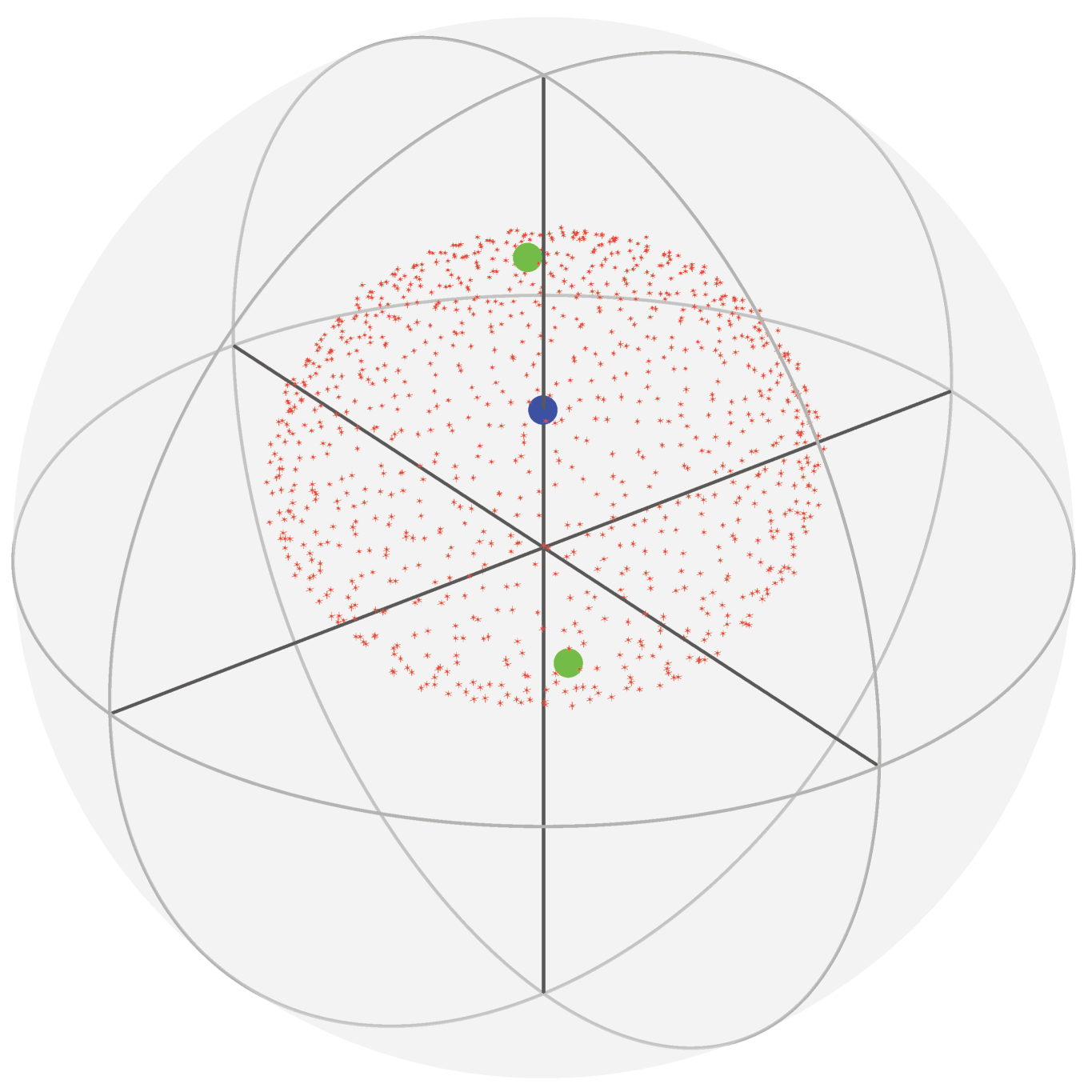}	&\includegraphics[width=1in]{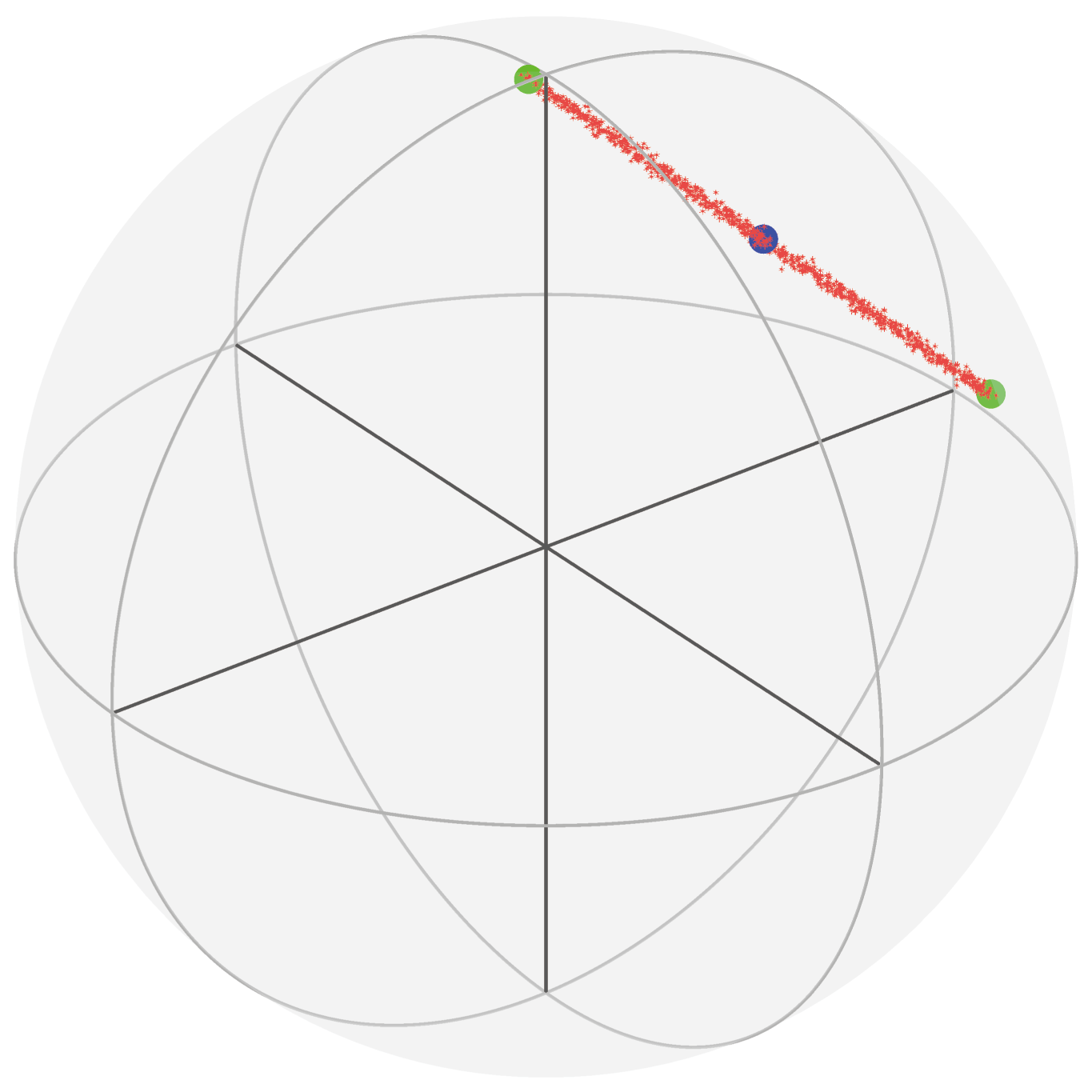}	&\includegraphics[width=1in]{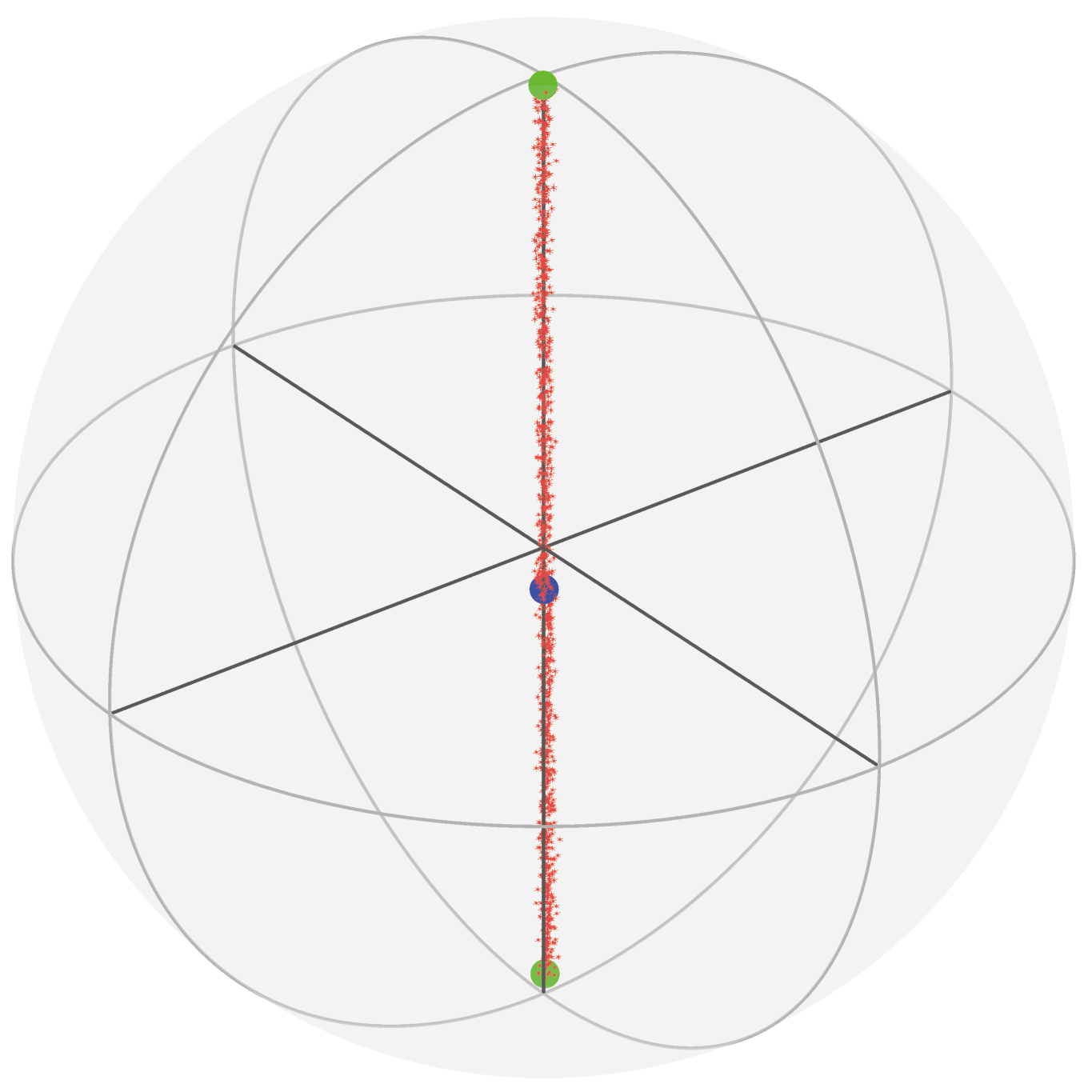}	&	\includegraphics[width=1in]{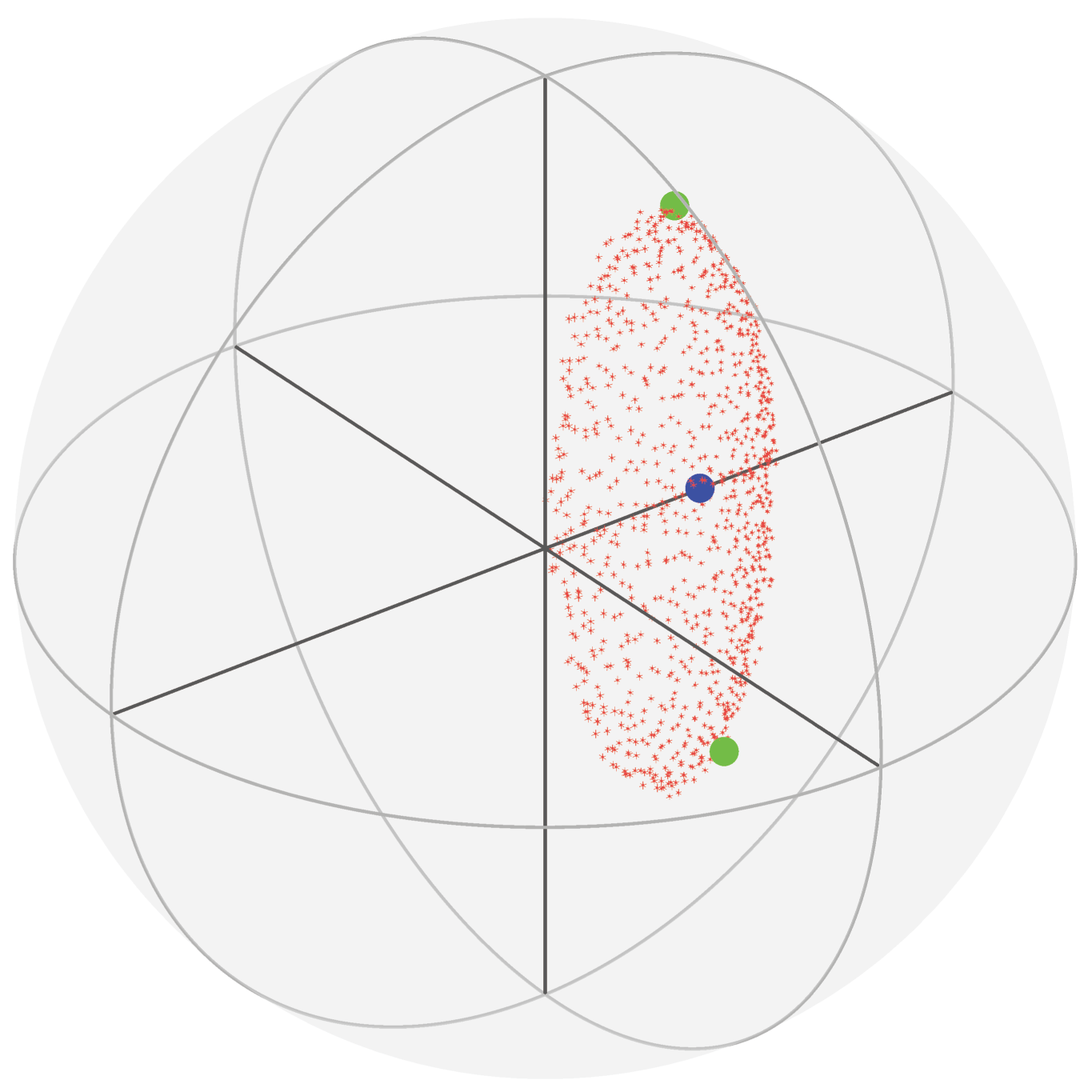}&\includegraphics[width=1in]{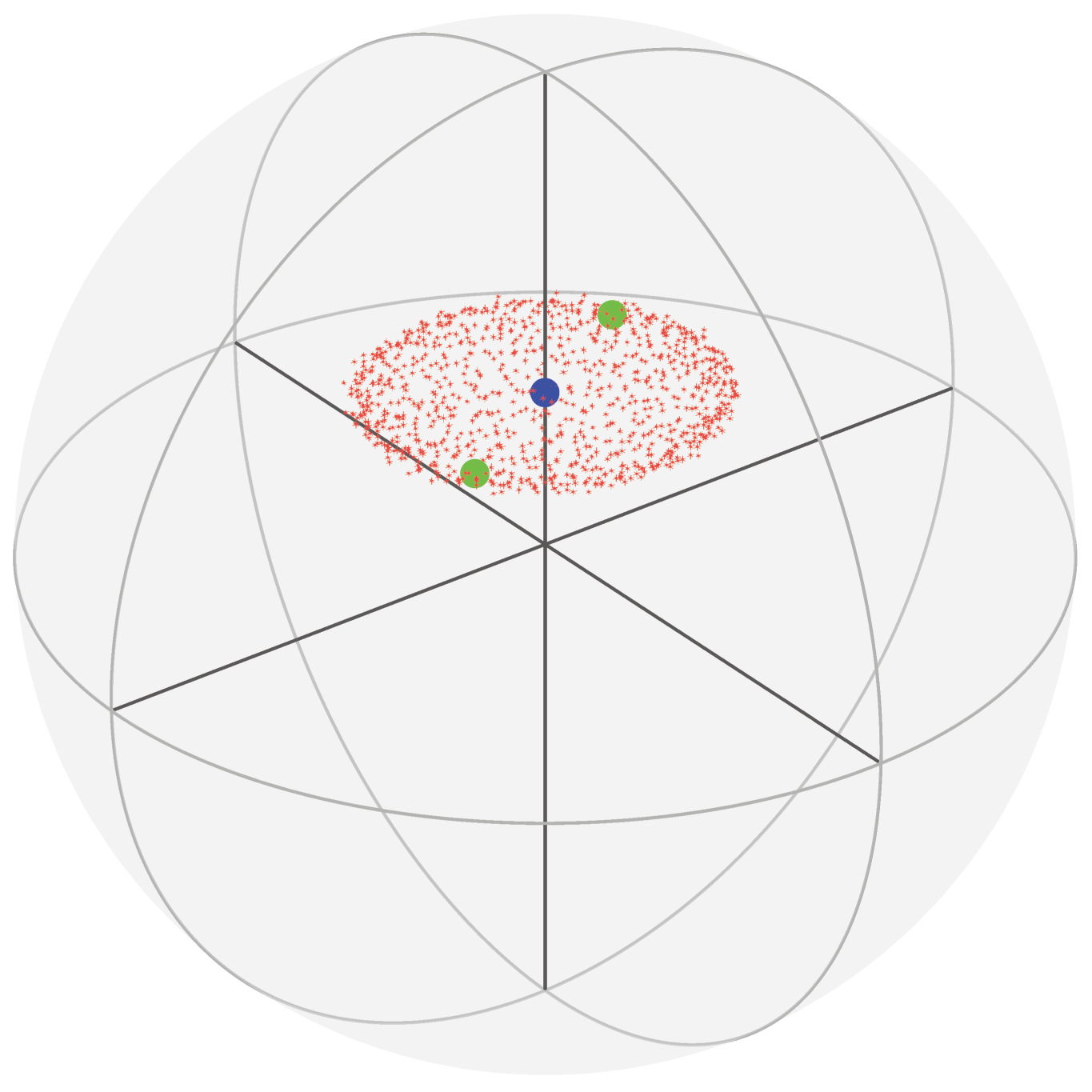}\\
				$\mathcal{E}_{B|A}$	& \includegraphics[width=1in]{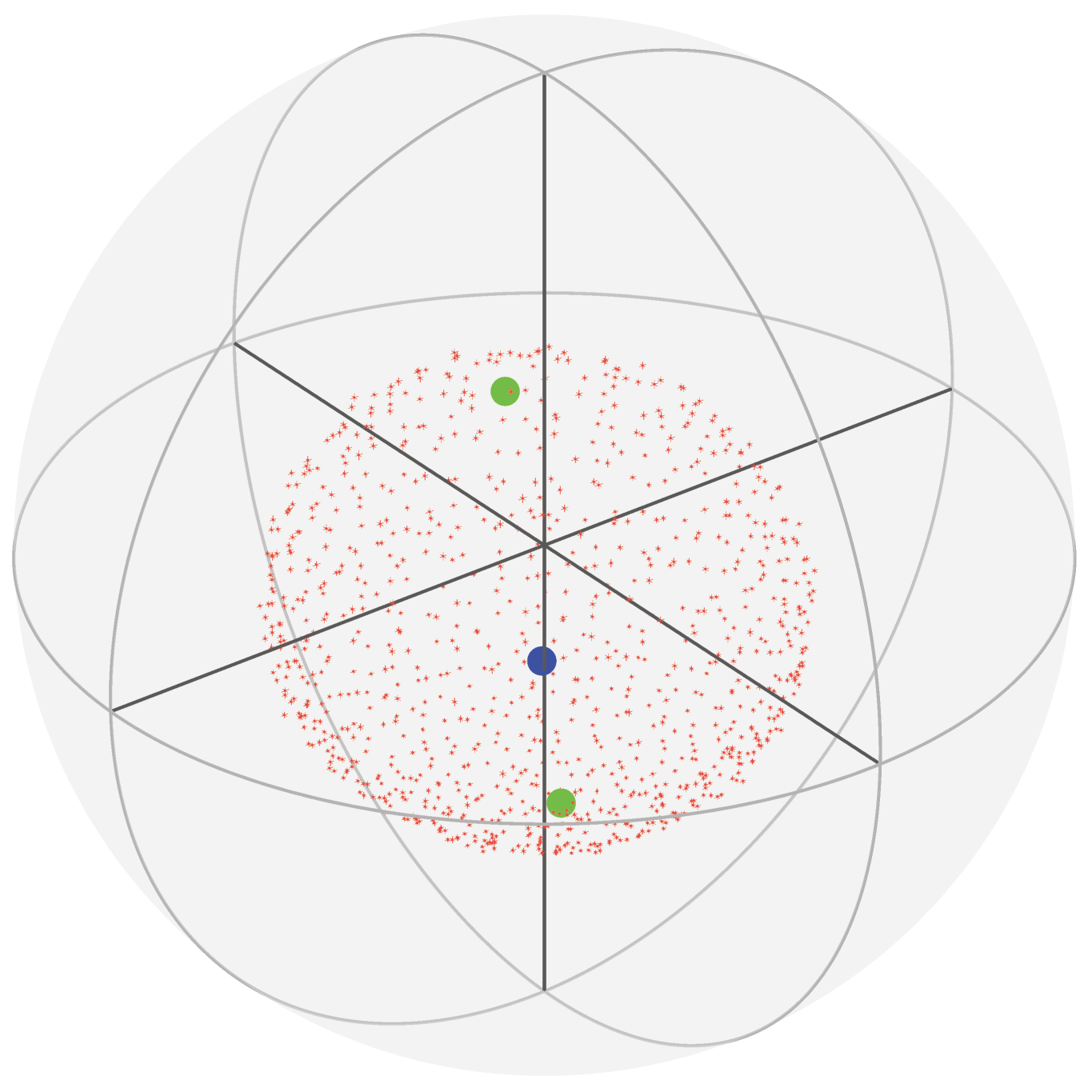} &\includegraphics[width=1in]{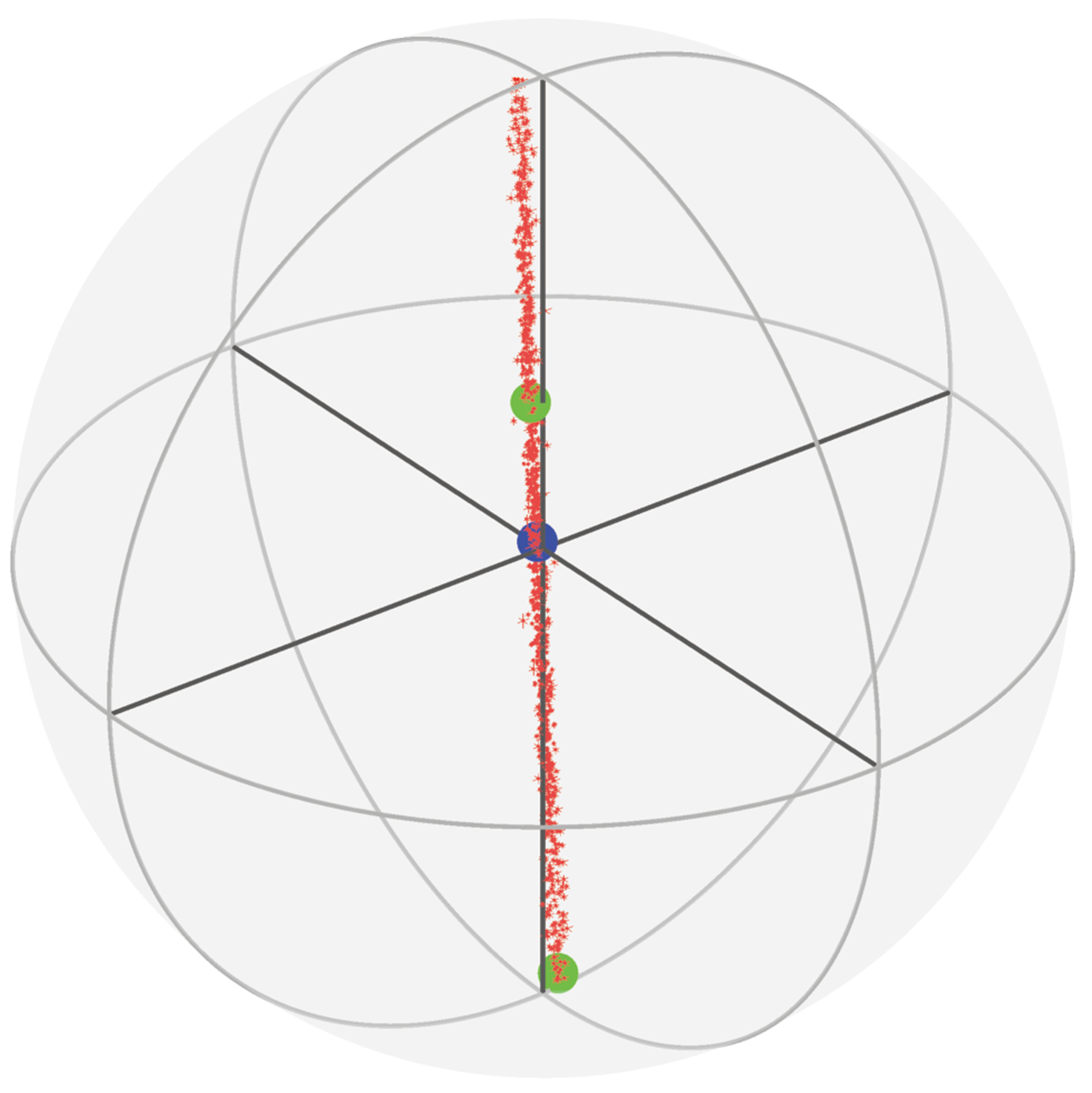} &  \includegraphics[width=1in]{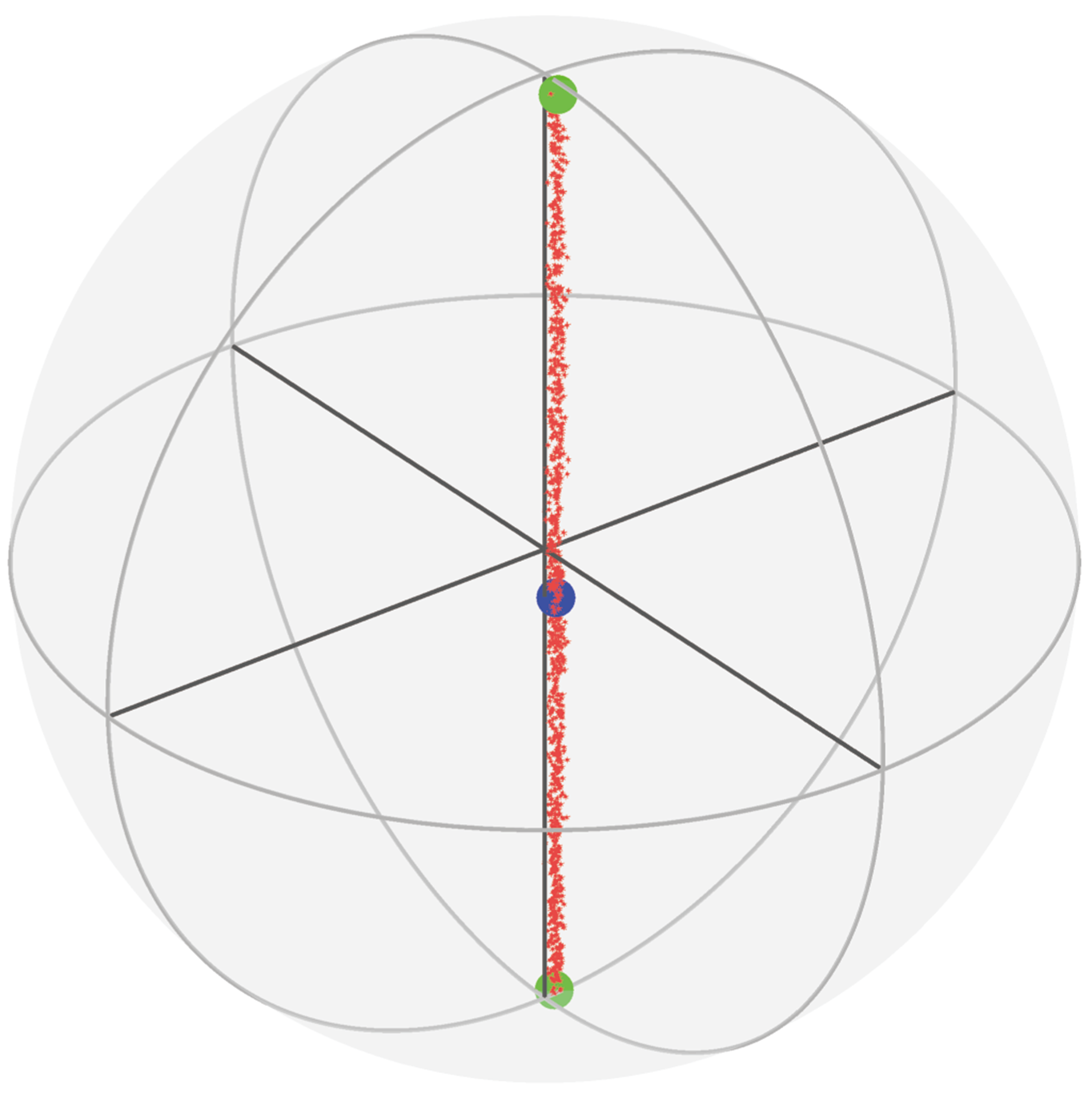}	&\includegraphics[width=1in]{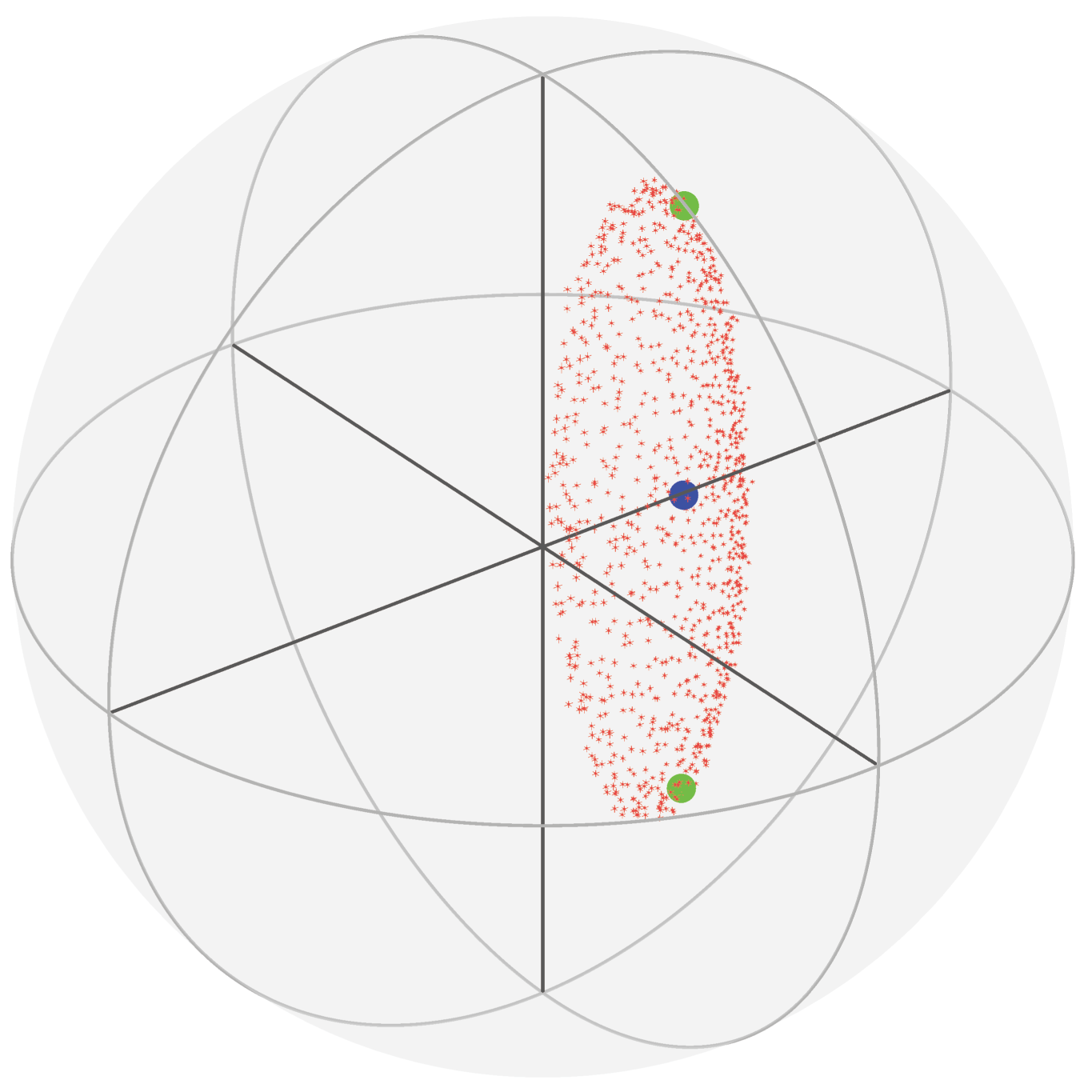}&\includegraphics[width=1in]{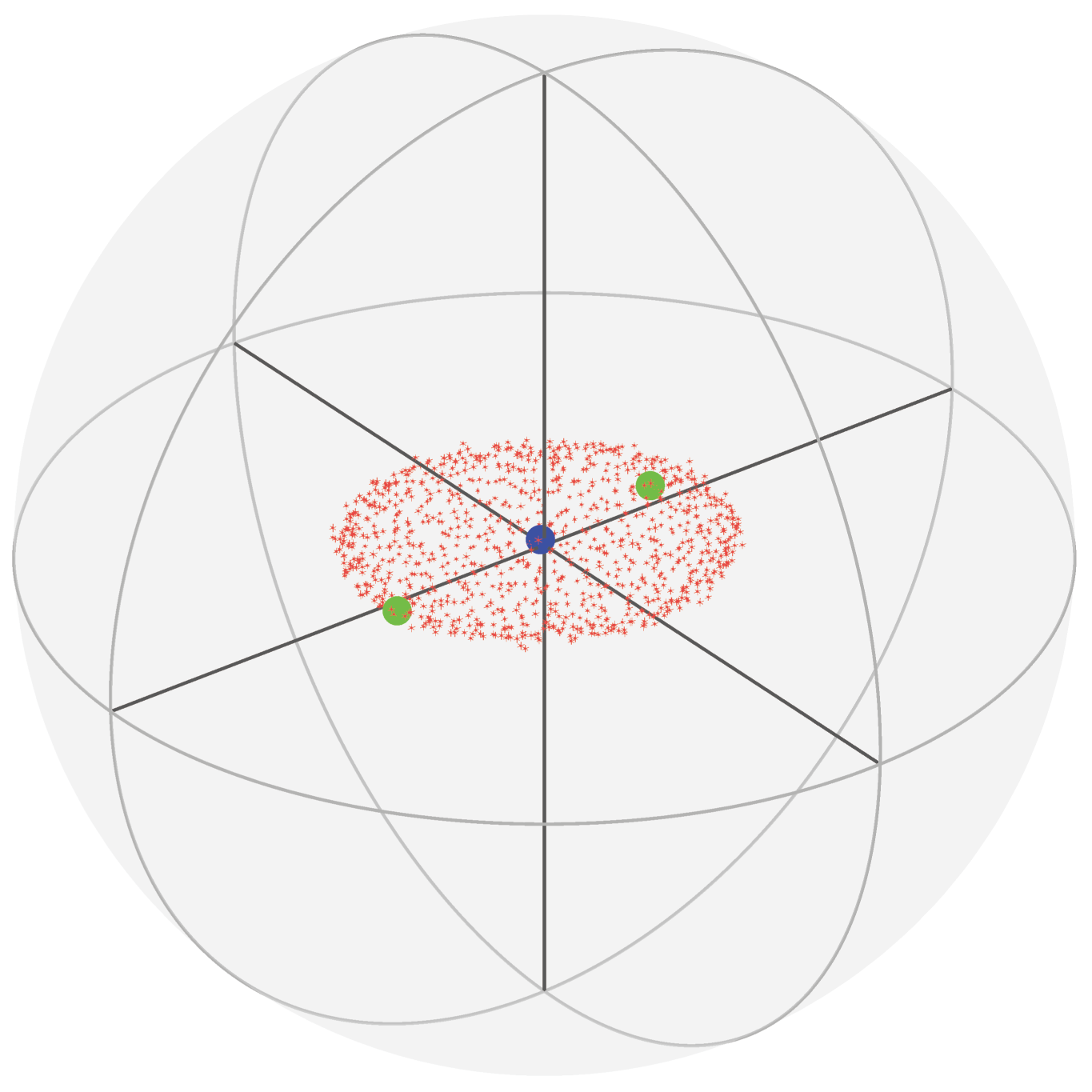}\\
				Con. & 0.2395$\pm 0.0014$ (0.2500) & 0(0)&0.0020$\pm 0.0014$(0)  &0(0)&0(0) \\
			\end{tabular}
			\end{ruledtabular}		
	\end{center}
					\caption{\textbf{The zoo of quantum steering ellipsoids}. Both Alice's (Bob's) steering ellipsoid $\mathcal{E}_{B|A}$ ($\mathcal{E}_{A|B}$) are constructed for all states $\rho_1$-$\rho_8$ in Table~\ref{tab_states}. Additionally, the degree of entanglement, measured by concurrence, is obtained for these experimentally reconstructed states, while the theoretical values are given in parentheses. In each quantum steering ellipsoid, the blue dot describes the reduced state $\rho_A$ or $\rho_B$, and two green points represent the corresponding steered states by one single measurement $\{E_1=\ket{0}\bra{0}, E_2=\ket{1}\bra{1}\}$, except for $\rho_8$ with two orthogonal projectors ${\psi_1}=\cos3\pi/16\ket{0}+e^{i\pi/10}\sin3\pi/16\ket{1}$ and ${\psi_2}=\sin3\pi/16\ket{0}-e^{i\pi/10}\cos3\pi/16\ket{1}$. Abbreviations in this figure are: Sep., Separable; Ent., Entangled; Incomp., Incomplete steering; Comp., Complete steering; Con., Concurrence.}\label{tab_zoo}
		\end{figure*}

			\section{Experimental results}\label{results}
			
			By preparing states in the degree of polarization and path, we obtain all two-qubit states in Table~\ref{tab_states} with nearly perfect fidelity and high generation rate. We first use quantum state tomography to estimate the prepared state $\rho^{\rm exp}$ from experimental data and leave the detailed data analysis in~\cite{SM}. In our experiments, an average fidelity of $0.9926 \pm 0.0087$ is achieved for these states~\cite{SM}, where the state fidelity is calculated by $F=\left(\tr{\rt{\rt{\rho^{\rm exp}}\rho\rt{\rho^{\rm exp}}}}\right)^2$~\cite{Jozsa1994}. We then construct the steering ellipsoids, including the degenerate pancake and needle, to test whether they are entangled, EPR-steerable, discord, and complete steering. Each  steered state is reconstructed from $5.0 \times 10^4$ detection events via quantum state tomography, and the corresponding steering ellipsoid, plotted in Fig.~\ref{tab_zoo}, is constructed via $1000$ measurement points.

			As shown in Fig.~\ref{tab_zoo}, the steering ellipsoid that coincides with the whole Bloch ball is constructed for the partially entangled pure state $\rho_1$, and the QSEs corresponding to a series of Werner states with $p_1=1/2, p_2=1/3$, and $p_3=1/5$, are constructed to further test the nested tetrahedron condition for entanglement. It is observed that for each $p$, the steering ellipsoid is nearly centered at the origin with three semiaxes approximately close to $p$~\cite{SM}, and also confirms that $p=1/3$ is the boundary between separability and entanglement for Werner states because the largest sphere that can be inscribed inside a tetrahedron inside the unit sphere has a radius less than $1/3$.  Furthermore, the ellipsoid for $\rho_6$ becomes a
			a segment of a diameter, thus having zero discord. 
			
			To reveal the property of steerability, we first use $\rho_3$ to construct Alice's steering ellipsoid that Bob can never have EPR-steerability of Alice and Bob's ellipsoid that Alice is able to steer Bob. With respect to the steering completeness which requires there being a measurement for Alice steer Bob to any set of states generated from the state decomposition of Bob's local state, it is shown in Fig.~\ref{tab_zoo} via $\rho_5$ and $\rho_8$ that it is not the case. It is in particular found that   the measurement direction $E_1=|0\rangle\langle0|$ by Alice steers Bob to one end  (green point in Fig.~\ref{tab_zoo}) of the needle $\mathcal{E}_{B|A}$ while the complementary measurement direction $E_2=|1\rangle\langle1|$ does not steer Bob to the other end, however, the reduced state $\rho_B$ always admits such a state decomposition of which two states are located on the surface of the steering ellipsoid. This immediately certifies the presence of the steering incompleteness, which is also confirmed via $\rho_8$ with $\mathcal{E}_{B|A}$.

			\begin{figure}[htb]
				\begin{tabular}{|M|M|M|M|M|}
					Shape& State& Setting &$\mathcal{E}_{A|B}$&	$\mathcal{E}_{B|A}$\\
					\hline
					Ellipsoid& $\rho_4$	&	\includegraphics[width=0.7in]{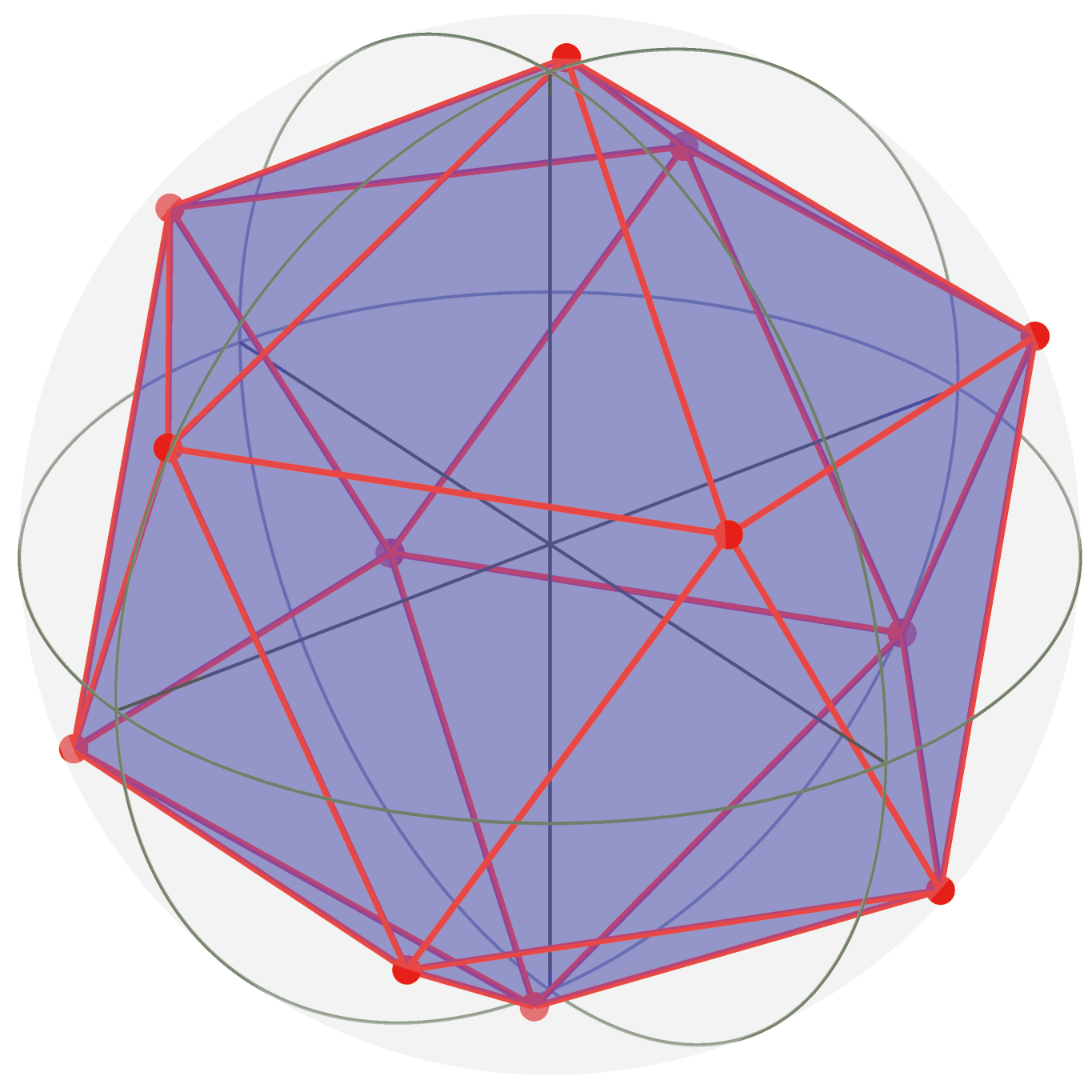}	&	\includegraphics[width=0.7in]{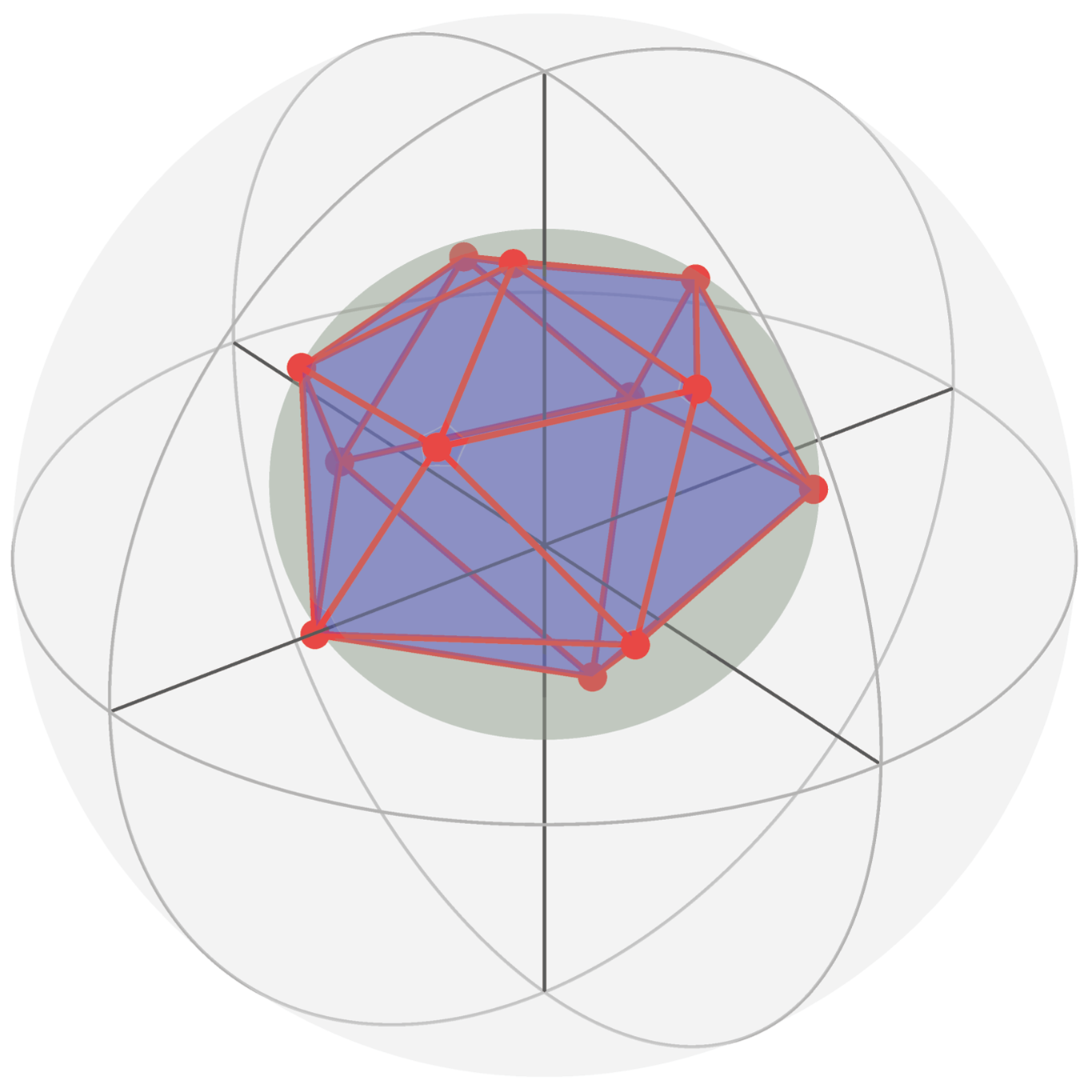}	&	\includegraphics[width=0.7in]{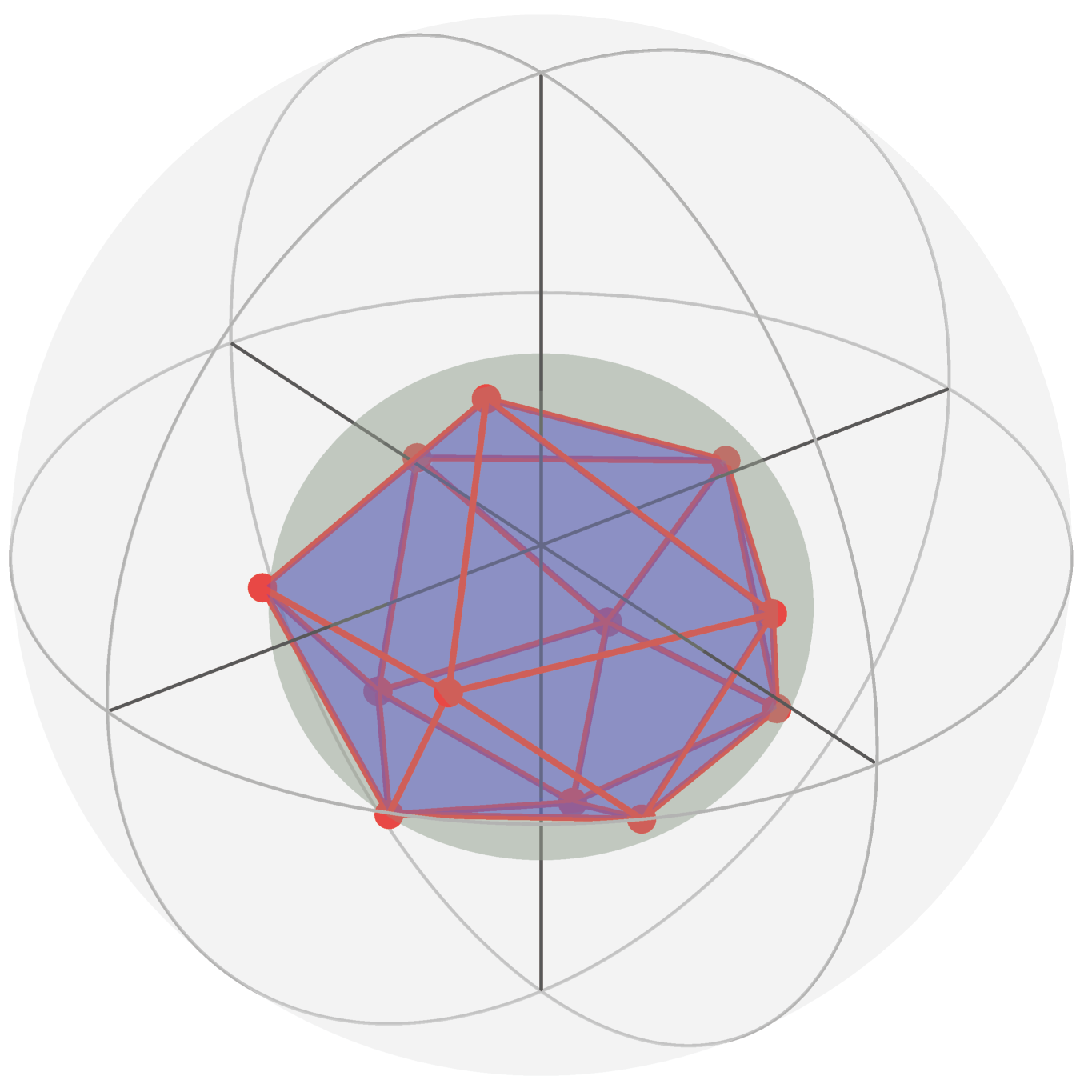}\\
					Ellipse& $\rho_8$	&	\includegraphics[width=0.7in]{icosahedron-1.png}	&	\includegraphics[width=0.7in]{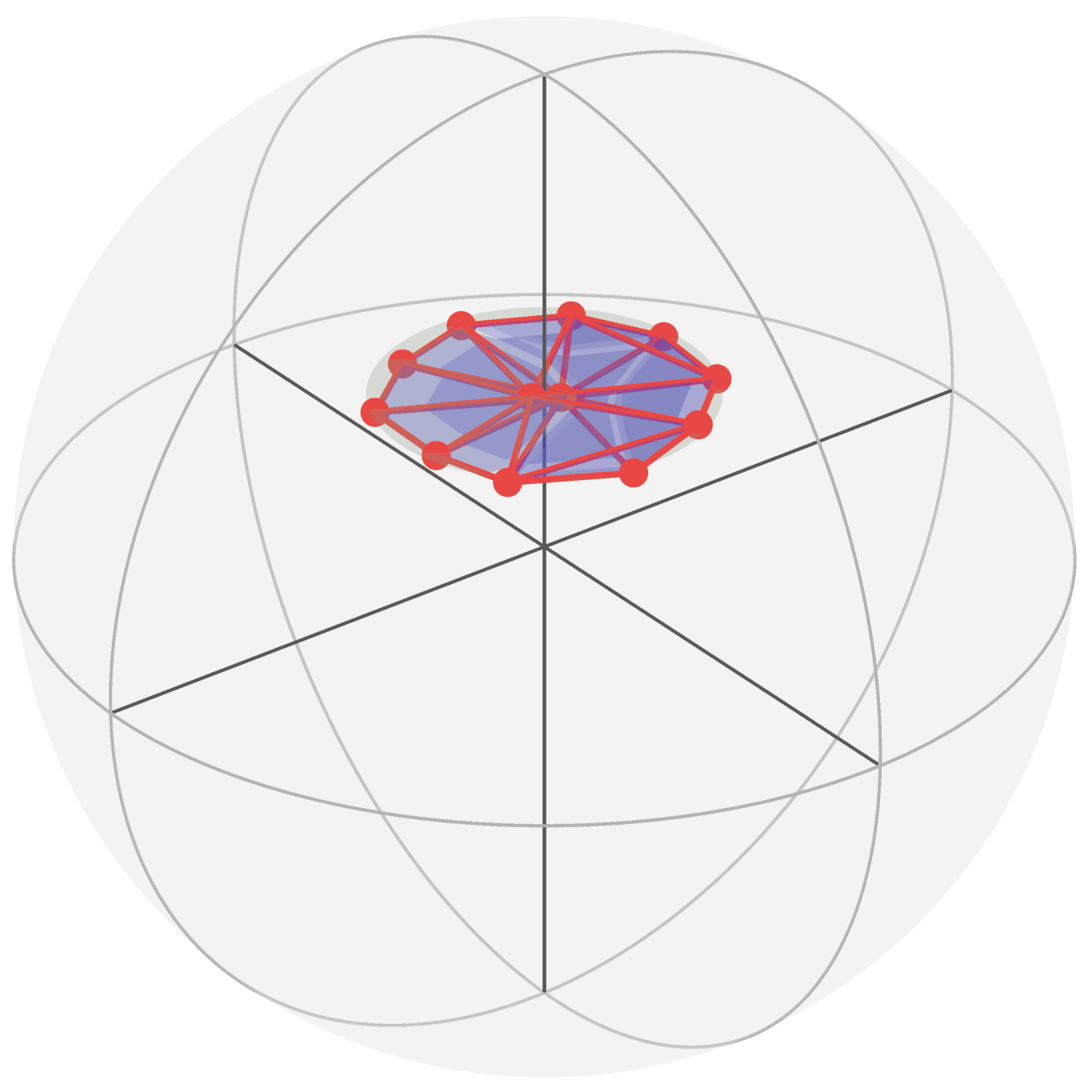}	&	\includegraphics[width=0.7in]{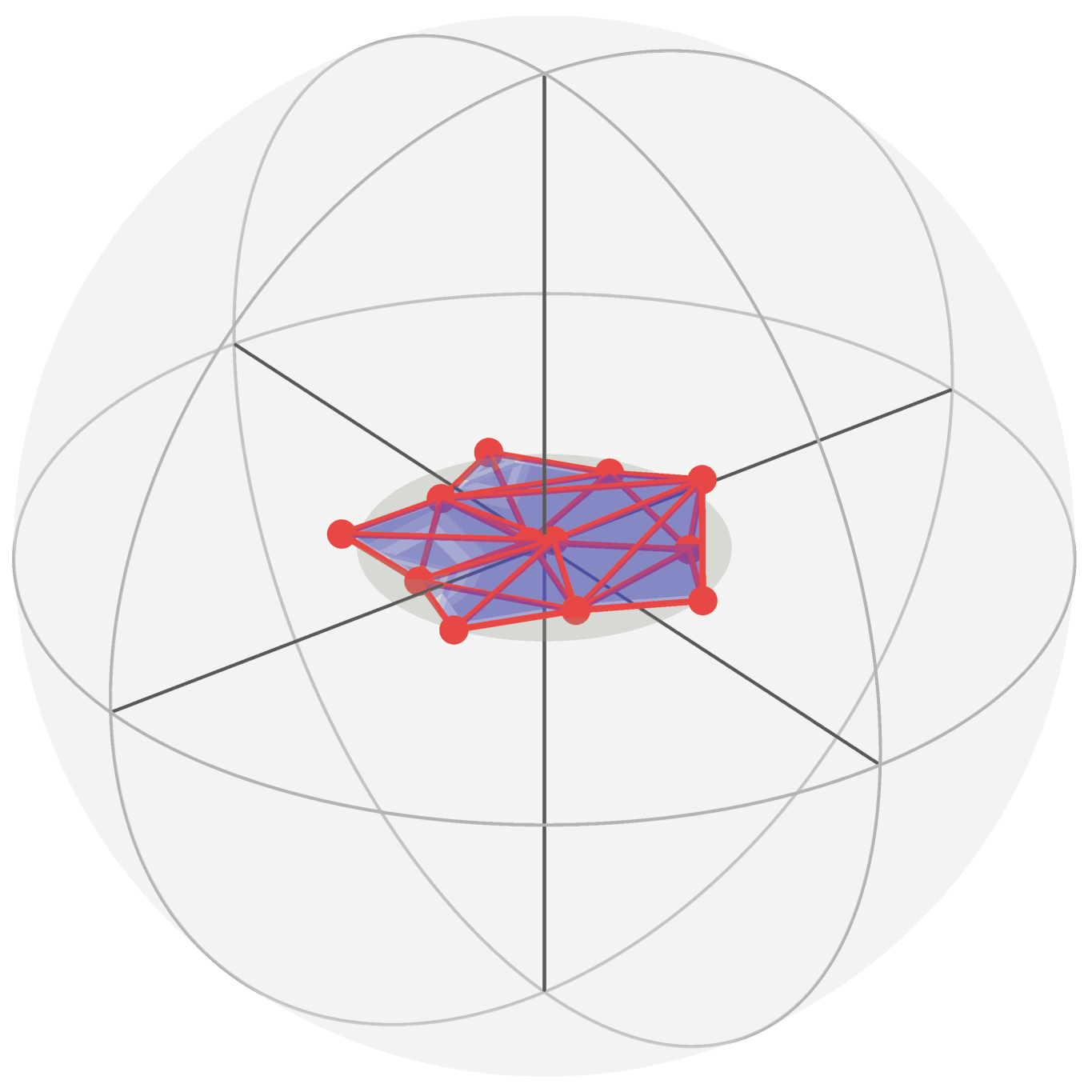}\\
					Line& $\rho_6$	&	\includegraphics[width=0.7in]{icosahedron-1.png}	&	\includegraphics[width=0.7in]{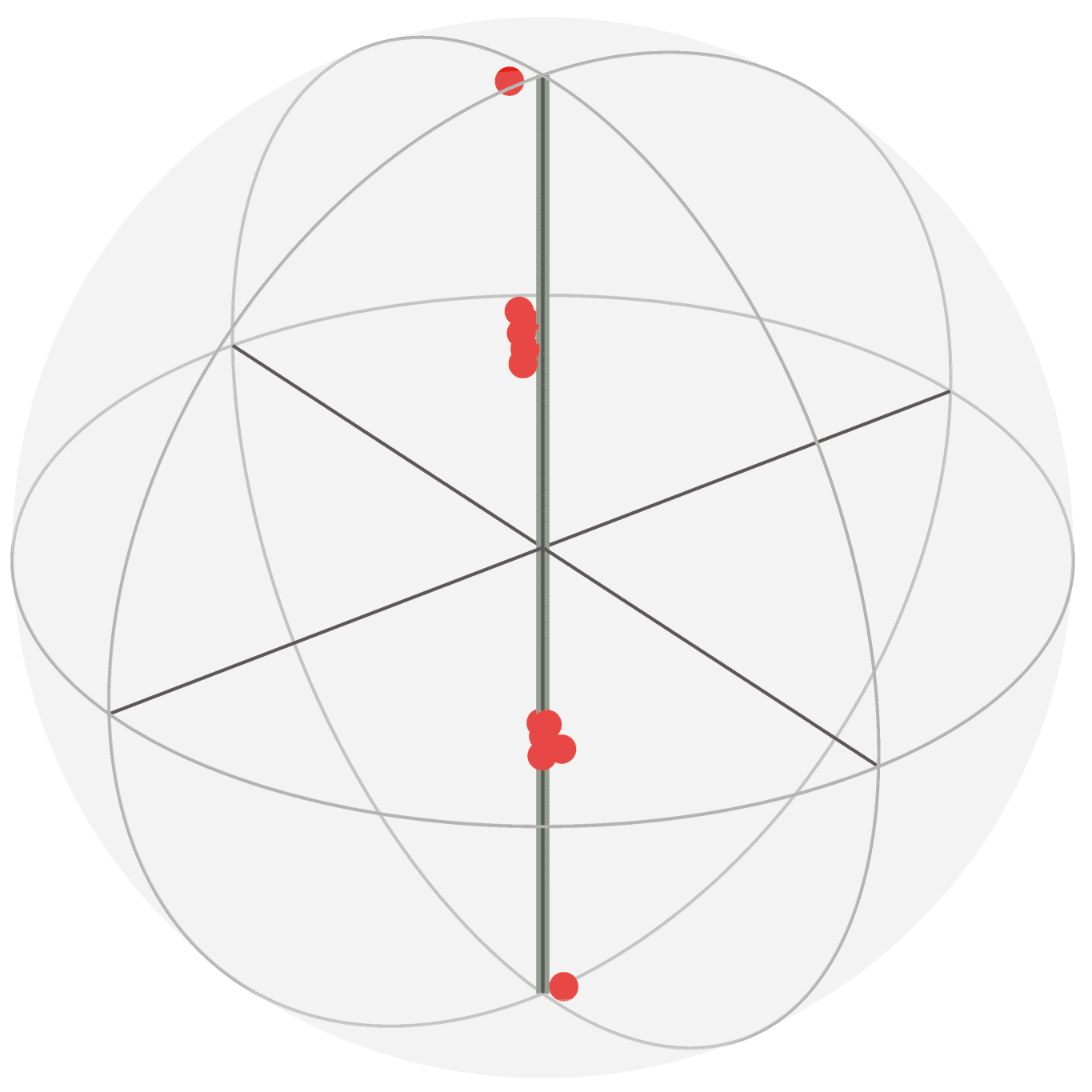}	&	\includegraphics[width=0.7in]{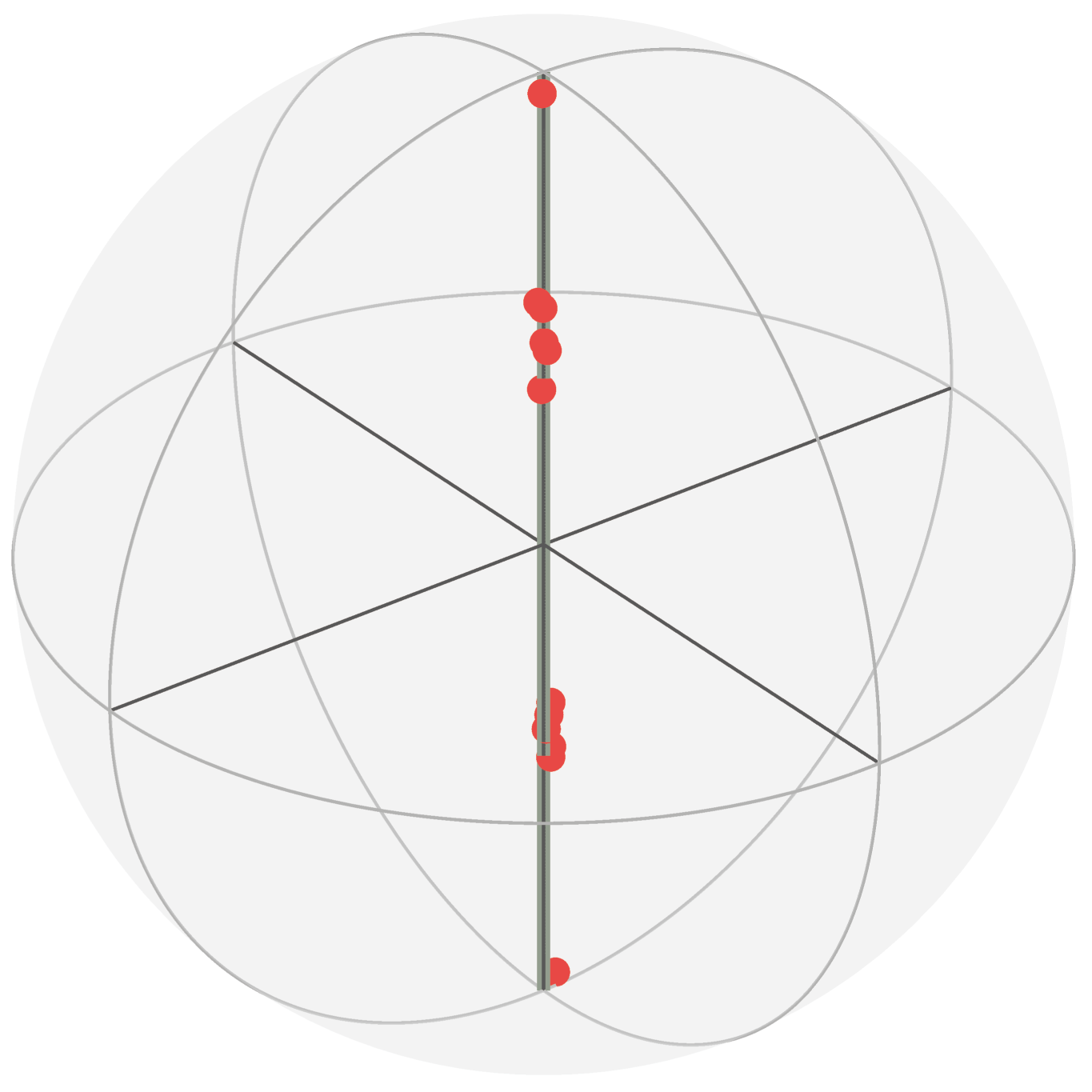}
				\end{tabular}
				\caption{\textbf{Twelve points are enough to determine the quantum steering ellipsoid}. The twelve vertices of an icosahedron are choices for measurement directions on three states, and the corresponding QSEs are reconstructed.  }\label{re_little}
			\end{figure}

			We also present an efficient approach to reconstruct QSEs for two-qubit states. Following directly from Eqs.~(\ref{center}) and~(\ref{orientation}) that nine points generically determine an ellipsoid, we choose the twelve vertices of an icosahedron as measurement directions to construct the steering ellipsoid, instead of running all possible directions. Our method is implemented on the states $\rho_4,\rho_8$, and $\rho_6$, of which the steering ellipsoids are an ellipsoid, ellipse, and straight line.  As displayed in Fig.~\ref{re_little}, the red dots represent twelve data points sampled from the vertices of an icosahedron and are fitted as an ellipsoid, an ellipse, or a line, while the dark gray area describes the QSE predicted by theory. In order to test the robustness of our method, we further perform 50 experiments by randomly rotating the icosahedron and thus its vertices, and calculate the geometric properties of QSEs, i.e., the volume for $\rho_4$, area for $\rho_8$, and length for $\rho_6$. It is found that it is able to fit the predicted ellipsoids with high precision~\cite{SM}.

			\section{Discussion and Conclusion}
			
			We have experimentally verified the zoo of quantum steering ellipsoids by generating eight different states. It is found that the QES can not only provide a faithful geometric characterization of the shared two-qubit state, but also reflects almost all nonclassical features in its geometric properties, such as entanglement, EPR-steering, discord, and steering incompleteness. 
			
			It will be interesting to apply the QES to reveal other nonclassical features of the qubit system, such as the steered coherence~\cite{Hu2015} and measurement reality~\cite{Hall2019}. It is also of both theoretical and experimental interest to investigate the quantum steering ellipsoid in multiple spins to observe quantum phase transitions~\cite{Du2021,Li2023,Rosario2023} and in higher dimensional system beyond qubits. Moreover, the efficient approach of using  the vertices of an icosahedron to construct the steering ellipsoids is expected to be a potential tool in future quantum networks to characterize quantum correlations without shared reference frames among distant parties \cite{Wollman2018}.

			\section*{Acknowledgement}
			This work is supported by the Shanghai Municipal Science and Technology Fundamental Project (No. 21JC1405400), the Fundamental Research Funds for the Central Universities (Nos. 22120230035, WK2030000061, YD2030002015), the National Natural Science Foundation of China (Nos. 12205219 and 62075208), the innovation Program for Quantum Science and Technology (No. 2021ZD0301604). This work was partially carried out at the USTC Center for Micro and Nanoscale Research and Fabrication.


\begin{thebibliography}{99}
			
			\bibitem{Schrodinger1935}
			E. Schrödinger, Discussion of Probability Relations between Separated Systems, \href{https://www.cambridge.org/core/journals/mathematical-proceedings-of-the-cambridge-philosophical-society/article/discussion-of-probability-relations-between-separated-systems/C1C71E1AA5BA56EBE6588AAACB9A222D}{Proc. Cambridge Philos. Soc. \textbf{31}, 555 (1935).}
			
			\bibitem{Schrodinger1936}
			E. Schrödinger, Probability relations between separated systems, \href{https://www.cambridge.org/core/journals/mathematical-proceedings-of-the-cambridge-philosophical-society/article/probability-relations-between-separated-systems/641DDDED6FB033A1B190B458E0D02F22}{Proc. Cambridge Philos. Soc. \textbf{32}, 446 (1936).}
			
			\bibitem{Hughston1993}
			L. P. Hughston, R. Jozsa, and W. K. Wootters, A complete classification of quantum ensembles having a given density matrix, \href{https://www.sciencedirect.com/science/article/abs/pii/0375960193908809?via%3Dihub}{Phys. Lett. A \textbf{183}, 14 (1993).}
				
			\bibitem{Gisin1996}
			N. Gisin, Hidden quantum nonlocality revealed by local filters, \href{https://www.sciencedirect.com/science/article/pii/S0375960196800016?via%3Dihub}{Phys. Lett. A \textbf{210}, 151 (1996).}
			
			\bibitem{Verstraete2002}
			F. Verstraete, Ph.D. thesis, Katholieke Universiteit Leuven (2002). 
			
			\bibitem{Wiseman2007}
			H. M. Wiseman, S. J. Jones, and A. C. Doherty, Steering, Entanglement, Nonlocality, and the Einstein-Podolsky-Rosen Paradox, \href{https://journals.aps.org/prl/abstract/10.1103/PhysRevLett.98.140402}{Phys. Rev. Lett. \textbf{98}, 140402 (2007).}
			
			\bibitem{Sania2014}
			S. Jevtic, M. Pusey, D. Jennings, and T. Rudolph, Quantum Steering Ellipsoids, \href{https://journals.aps.org/prl/abstract/10.1103/PhysRevLett.113.020402}{ Phys. Rev. Lett. \textbf{113}, 020402 (2014).} 
			
				\bibitem{Milne2014B}
			A. Milne, D. Jennings, S. Jevtic, and T. Rudolph, Quantum correlations of two-qubit states with one maximally mixed marginal, \href{https://journals.aps.org/pra/abstract/10.1103/PhysRevA.90.024302}{Phys. Rev. A \textbf{90}, 024302 (2014).}
			
				\bibitem{Jevtic2015}
			S. Jevtic, M. J.W. Hall, M. R. Anderson, M. Zwierz, and H. M. Wiseman, Einstein–Podolsky–Rosen steering and the steering ellipsoid,\href{https://opg.optica.org/josab/fulltext.cfm?uri=josab-32-4-A40&id=314132}{ J. Opt. Soc. Am. B \textbf{32}, A40 (2015).}
			
			\bibitem{Chau2016}
			H. C. Nguyen and T. Vu, Nonseparability and steerability of two-qubit states from the geometry of steering outcomes, \href{https://journals.aps.org/pra/abstract/10.1103/PhysRevA.94.012114}{Phys. Rev. A \textbf{94}, 012114 (2016).}
			
				\bibitem{Chau2016B}
			H. Chau Nguyen and T. Vu, Necessary and sufficient condition for steerability of two-qubit states by the geometry of steering outcomes, \href{https://iopscience.iop.org/article/10.1209/0295-5075/115/10003}{Europhys. Lett. \textbf{115}, 10003 (2016).}
			
				\bibitem{Quan2016}
			Q. Quan, H. Zhu, S.-Y. Liu, S.-M. Fei, H. Fan, and W.-L. Yang, Steering Bell-diagonal states, \href{https://www.nature.com/articles/srep22025}{Sci. Rep. \textbf{6}, 22025 (2016).}
			
			\bibitem{McClosky2017}
			R. McCloskey, A. Ferraro, and M. Paternostro, Einstein-Podolsky-Rosen steering and quantum steering ellipsoids: Optimal two-qubit states and projective measurements, \href{https://journals.aps.org/pra/abstract/10.1103/PhysRevA.95.012320}{Phys. Rev. A \textbf{95},	012320 (2017).}
			
			\bibitem{Chau2017}
			H. C. Nguyen and K. Luoma, Pure steered states of Einstein-Podolsky-Rosen steering, \href{https://journals.aps.org/pra/abstract/10.1103/PhysRevA.95.042117}{Phys. Rev. A \textbf{95}, 042117 (2017).}
			
				\bibitem{Song2023}
			Q.-C. Song, T. Baker, and H. M. Wiseman, On the power of one pure steered state for EPR-steering with a pair of qubits, \href{https://iopscience.iop.org/article/10.1088/1367-2630/accfba}{New J. Phys. \textbf{25}, 053005 (2023).} 
			
				\bibitem{Milne2014}
			A. Milne, S. Jevtic, D. Jennings, H. Wiseman, and T. Rudolph, Quantum steering ellipsoids, extremal physical states and monogamy, \href{https://iopscience.iop.org/article/10.1088/1367-2630/16/8/083017}{New J. Phys. \textbf{16}, 083017 (2014).} 
			
			\bibitem{Shi2011}
			M. Shi, F. Jiang, C. Sun, and J. Du, Geometric picture of quantum discord for two-qubit quantum states, \href{https://iopscience.iop.org/article/10.1088/1367-2630/13/7/073016}{New J. Phys. \textbf{13}, 073016 (2011).}
			
			\bibitem{Shi2012}
			M. Shi, C. Sun, F. Jiang, X. Yan, and J. Du, Optimal measurement for quantum discord of two-qubit states, \href{https://journals.aps.org/pra/abstract/10.1103/PhysRevA.85.064104}{Phys. Rev. A \textbf{85}, 064104 (2012).}
		
			\bibitem{Hu2015}
			X. Hu and H. Fan, Effect of local channels on quantum steering ellipsoids, \href{https://journals.aps.org/pra/abstract/10.1103/PhysRevA.91.022301}{Phys. Rev. A \textbf{91}, 022301 (2015).}
			
			\bibitem{Hu2016}
			X. Hu, A. Milne, B. Zhang, and H. Fan, Quantum coherence of steered states, \href{https://www.nature.com/articles/srep19365}{Sci. Rep. \textbf{6}, 19365 (2016).}
			
			\bibitem{Hall2019}
			M. J. W. Hall and \'A. Rivas, Geometry of joint reality: Device-independent steering and operational completeness, \href{https://journals.aps.org/pra/abstract/10.1103/PhysRevA.100.062105}{Phys. Rev. A \textbf{100}, 062105 (2019).}
			

			\bibitem{Du2021}
			M.-M. Du, D.-J. Zhang, Z.-Y. Zhou, and D. M. Tong, Visualizing quantum phase transitions in the XXZ model via the quantum steering ellipsoid,\href{https://journals.aps.org/pra/abstract/10.1103/PhysRevA.104.012418}{Phys. Rev. A \textbf{104}, 012418 (2021).}
			
			\bibitem{Li2023}
			C.-X. Li, S. Yang, J.-B. Xu, and H.-Q. Lin, Exploring dynamical quantum phase transitions in a spin model with deconfined critical point via the quantum steering ellipsoid, \href{https://journals.aps.org/prb/abstract/10.1103/PhysRevB.107.085130}{Phys. Rev. B \textbf{107}, 085130 (2023). } 
			
			\bibitem{Rosario2023}
			P. Rosario, A. C. Santos, Quantum steering ellipsoids and quantum obesity in critical systems, \href{https://doi.org/10.48550/arXiv.2312.12537}{arXiv.2312.12537 (2023)}
			

			\bibitem{Cheng2016}
			S. Cheng, A. Milne, M. J.W. Hall, and H. M. Wiseman, Volume monogamy of quantum steering ellipsoids for multiqubit systems, \href{https://journals.aps.org/pra/abstract/10.1103/PhysRevA.94.042105}{Phys. Rev. A \textbf{94}, 042105 (2016).}
			
			\bibitem{Zhang2019}
			C. Zhang, S. Cheng, L. Li, Q.-Y. Liang, B.-H. Liu, Y.-F. Huang, C.-F. Li, G.-C. Guo, M. J. W. Hall, H. M. Wiseman, and G. J. Pryde, Experimental Validation of Quantum Steering Ellipsoids and Tests of Volume Monogamy Relations, \href{https://journals.aps.org/prl/abstract/10.1103/PhysRevLett.122.070402}{Phys. Rev. Lett. \textbf{122}, 070402 (2019).}
			
			\bibitem{Div2023}
			B. G. Divyamani, I. Reena, Prasanta K. Panigrahi, A. R. Usha Devi, and Sudha, Canonical steering ellipsoids of pure symmetric multiqubit states with two distinct spinors and volume monogamy of steering, \href{https://journals.aps.org/pra/abstract/10.1103/PhysRevA.107.042207}{Phys. Rev. A \textbf{107}, 042207 (2023).}
			
			\bibitem{Song2023B}
			Q.-C. Song, T. Baker, and H. M. Wiseman, Shareability of steering in 2-producible states, \href{https://journals.aps.org/pra/abstract/10.1103/PhysRevA.108.012216}{Phys. Rev. A \textbf{108}, 012216 (2023).} 
			
			\bibitem{Cheng2018}
			S. Cheng, \href{https://research-repository.griffith.edu.au/bitstream/handle/10072/380058/Cheng,%20Shuming_Final%20Thesis_redacted.pdf?sequence=1}{Ph.D. thesis, Griffith University (2018).} 
			
			\bibitem{Werner1989}
			R. F. Werner, Quantum states with Einstein-Podolsky-Rosen correlations admitting a hidden-variable model, \href{https://journals.aps.org/pra/abstract/10.1103/PhysRevA.40.4277}{Phys. Rev. A \textbf{40}, 4277 (1989).}
			

			
			\bibitem{Zurek2001}
			H. Ollivier and W. H. Zurek, Quantum Discord: A Measure of the Quantumness of Correlations, \href{https://journals.aps.org/prl/abstract/10.1103/PhysRevLett.88.017901}{Phys. Rev. Lett. \textbf{88}, 017901 (2001). }
			
			\bibitem{Henderson2001}
			L. Henderson and V. Vedral, Classical, quantum and total correlations,\href{https://iopscience.iop.org/article/10.1088/0305-4470/34/35/315}{J. Phys. A: Math. Gen. \textbf{34}, 6899 (2001).}
			
			\bibitem{Modi2012}
			K. Modi, A. Brodutch, H. Cable, T. Paterek, and V. Vedral, The classical-quantum boundary for correlations: Discord and related measures, \href{https://journals.aps.org/rmp/abstract/10.1103/RevModPhys.84.1655}{Rev. Mod. Phys. \textbf{84}, 1655 (2012).}

			
			\bibitem{Bowles2016}
			J. Bowles, F. Hirsch, M. T. Quintino, and N. Brunner, Sufficient criterion for guaranteeing that a two-qubit state is unsteerable, \href{https://journals.aps.org/pra/abstract/10.1103/PhysRevA.93.022121}{Phys. Rev. A \textbf{93}, 022121 (2016).}
			
			\bibitem{SM}
			See supplementary material for detailed information.
			
		
			\bibitem{Zhang2015}
			C. Zhang, Y.-F. Huang, Z. Wang, B.-H. Liu, C.-F. Li, and G.-C. Guo, Experimental greenberger-horne-zeilinger-type six-photon quantum nonlocality, \href{https://journals.aps.org/prl/abstract/10.1103/PhysRevLett.115.260402}{ Phys. Rev. Lett. \textbf{115}, 260402 (2015).}
			
				
			
			\bibitem{Jozsa1994}
			R. Jozsa, Fidelity for Mixed Quantum States, \href{https://www.tandfonline.com/doi/abs/10.1080/09500349414552171}{J. Mod. Opt. \textbf{41}, 2315-2323 (1994).}
			
				\bibitem{Wollman2018}
			S. Wollmann, M. J. W. Hall, R. B. Patel, H. M. Wiseman, and G. J. Pryde, Reference-frame-independent Einstein-Podolsky-Rosen steering, \href{https://journals.aps.org/pra/abstract/10.1103/PhysRevA.98.022333}{Phys. Rev. A \textbf{98}, 022333 (2018).}
			
		
			
		
			\end{thebibliography}

			\setcounter{equation}{0}
			\renewcommand\theequation{S\arabic{equation}}
			
			\setcounter{table}{0}
			\renewcommand\thetable{S\arabic{table}}
			
			\setcounter{figure}{0}
			\renewcommand\thefigure{S\arabic{figure}}
			
			\clearpage
			
			\begin{widetext}
				
				\section*{Supplementary materials}

				\subsection{The zoo of nonclassical correlations}
				
				The state of a quantum system is described by a density matrix $\rho$ that is a nonnegative semidefinite operator with trace $1$.  It is essential to retrieve the state information via quantum measurement that is modelled as a positive-operator valued measure $\mathcal{M}=\{E_k\}$ with positive elements $M_i\succeq 0$ satisfying $\sum_k E_k=\id$. Ideally, performing a measurement $\mathcal{M}$ on state $\rho$ yields a probabilistic distribution, in the sense that each outcome $k$ associated with a measurement element $E_k$ happens with probability $p_k=\tr{\rho E_k}$.
				
				Suppose then that a bipartite quantum state $\rho_{AB}$ is shared by two parties, Alice and Bob say. The phenomenon of of {\it quantum steering}, first noticed by \sch~\cite{Schrodinger1935,Schrodinger1936}, describes that if Alice performs a measurement on her part, then Bob's system can be steered to a specific set of quantum states. Specifically, Alice's measurement outcome $k$ steers Bob's system to 
				\beq
				\rho^k_B=\frac{{\rm Tr}_A[\rho_{AB} E_k\otimes \mathbbm{I}]}{p_k} \label{steeredsta}
				\eeq
				with 
				\beq
				p_k=\tr{\rho_{AB} E_k\otimes \mathbbm{I}}. \label{steeredpro}
				\eeq
				Thus, collecting the Alice's full outcomes $k$ gives rise to a set of Bob's steered states $\{p_k, \rho^k_B\}$ satisfying
				\beq
				\sum _k p_k \rho^k_B =\rho_B. \label{descomposition}
				\eeq
				
				In particular, given a two-qubit state, quantum steering from Alice to Bob is fully captured by {\it quantum steering ellipsoid} (QSE) which visualizes the set of all Bob's possible states steered by  Alice performing all possible measurements in the Bloch picture~\cite{Sania2014}. Together with Alice’s and Bob’s local states, the QSE provides a faithful geometric representation of the shared two-qubit state, and thus generalizes the Bloch picture from the single qubit to two qubits. 
				
				Note further from Eq.~(\ref{descomposition}) that Alice's one single measurement on the shared $\rho_{AB}$ leads to a state-preparation process of Bob's local state $\rho_B$. Correspondingly, it is interesting to investigate whether any state decomposition of Bob's local state $\rho_B$ 
				\beq
				\rho_B= \sum_i p_i \rho^i_B \label{arbitrary}
				\eeq
				can be realized via the measurement process where Alice remotely measures a local measurement on the shared bipartite state $\rho_{AB}$. If it is possible for any given state $\rho_{AB}$, then there is a basic observation that any state $\rho_i$ in the decomposition~(\ref{arbitrary}) must be reachable via quantum steering as Eq.~(\ref{steeredsta}). With this observation, {\it completes steering} describes that given any realizable state decomposition of Bob's local state $\rho_B$~(\ref{arbitrary}), there always exists a measurement for Alice to steer Bob to the state set $\{p_i, \rho^i_B\}$ satisfying Eqs.~(\ref{steeredsta}) and~(\ref{steeredpro})~\cite{Sania2014}. Otherwise, it is called {\it incomplete steering}. It is explicitly shown in~\cite{Sania2014} that it is possible to witness incomplete steering in the two-qubit system, which is first experimentally confirmed in this work.
				
				If Alice measures a set of measurements $\mathcal{M}_j$, then Bob will receive the corresponding state assemblages $\{p_{k|j},\rho_B^{k|j}\}$.  {\it EPR-steering} is formulated as a task of entanglement verification~\cite{Wiseman2007} that amounts to checking if these states are prepared via a local hidden state (LHS) model in a form of
				\beq
				\rho_B^{k|j} =\sum_\lambda p(\lambda) \rho^B(\lambda) p(k|j, \lambda),
				\eeq
				where the hidden variable $\lambda$ specifies some classical probability distribution $p(k|j, \lambda)$ for Alice's measurement $j$ with outcome $k$. If there is no such LHS model, then EPR-steering from Alice to Bob is demonstrated.

				Finally, the definition of quantum discord is detailed. Entropy, as a measure of randomness or uncertainty in the system, is of fundamental and practical interest in statistical physics and information theory. Analogously, quantum entropy, measuring the information or uncertainty contained in the quantum system, also plays a crucial role in the field of quantum physics and quantum information. One notable example is the well-known von Neumann entropy
				\beq
				S(\rho):= \tr{\rho\log\rho}.
				\eeq
				Correspondingly, we are able to introduce the quantum mutual information
				\beq
				I(\rho_{AB}):= S(\rho_A)+S(\rho_B)-S(\rho_{AB})
				\eeq
				and the conditional quantum entropy
				\beq
				J_A(\rho_{AB}):=S(\rho_B)-S(\rho_B|\rho_A)  \label{conditionentropy}
				\eeq
				 for the bipartite state $\rho_{AB}$. The conditional entropy $J_A$~(\ref{conditionentropy}) depends on Alice's local measurement and thus represents the part of the correlations that can be attributed to classical correlations. Therefore,  it is possible to first maximize $J$ over the set of all possible (projective) measurements and then define quantum discord~\cite{Zurek2001}
				 \beq
				 \mathcal{D}_A(\rho_{AB}):=I(\rho_{AB})-\max_{\{E_k\}} J_{\{E_k\}} (\rho_{AB})=S(\rho_A)-S(\rho_{AB})+\min_{\{E_k\}} S(\rho|\{E_k\})
				 \eeq
			    to reveal the purely nonclassical correlations independently of measurement.
		
				\subsection{Experiment details}
				
			Here we show the detailed preparation process of two-qubit states listed in Table~\ref{tab_states} in the main text. Firstly, we use a type-\Rmnum{2} SPDC source to  generate the maximally entangled state:
				\begin{equation}
					\left | \phi_0  \right \rangle =(\left | HH \right \rangle +\left | VV \right \rangle )/\sqrt[]{2}.
				\end{equation}
		where H and V is horizontal and vertical polarisations of photon 1 and photon 2, which are labeled as 0 and 1 in the main text, respectively. Then photon pairs are injected into the optical paths through fibers, where the photon going through the up path is labelled as photon $1$ and the down path as photon $2$. The specific preparation process of states is shown in the light gray part of Fig.~\ref{setup} (\textbf{b})-(\textbf{i}).

		$\rho_1$ is a partial entangled state, which can be produced by placing a PPBS in the down paths to split an incident light beam with vertical polarization in a 50:50 ratio and make horizontal polarized light fully transparent(shown in Fig. \ref{setup}(\textbf{b})). We can get 
				\begin{equation}
					\rho_1 =|\psi_1\rangle \langle\psi_1|.
				\end{equation}
		where $\ket{\psi_1}=\cos\theta\ket{HH}+\sin\theta \ket{VV}$ and $\cos\theta=\rt{2/3}$.

		In the case of the Werner state, $\rho_2$ can be written as the mixture of two mixed states:
				\begin{equation}
					\begin{aligned}
						\rho_2 &= p\left|\psi_-\right\rangle\left\langle\psi_-\right| +\frac{1-p}{4}I_4\\
						&=	\frac{1-p}{2}\rho_{2a} + \frac{1+p}{2} \rho_{2b}.
					\end{aligned}\label{rho2}
				\end{equation} 
	where $\rho_{2a} = (|HH\rangle\langle HH| +|VV\rangle\langle VV|)/2$ and $\rho_{2b} = (|HV  \rangle\langle HV| +|VH \rangle\langle VH|+ a| HV\rangle\langle HV| +a| VH \rangle\langle VH|)/2 $ with $a = \frac{-2p}{1+p}$. The experimental states preparation is shown in Fig. \ref{setup}(\textbf{c}). Photon 1 does no operation. Photon 2 passes through the first BS and divides into two paths.  $\rho_{2a}$ is prepared by decoherencing the state $\phi_0$ completely with a thick quartz plate in the upper path , and $\rho_{2b}$ is prepared by an HWP rotated at $45^{\circ}$ and quartz plate in the lower path (Quartz plate is used to decoherence the quantum states with different level for three Werner states with $p = 1/2,1/3,1/5$ respectively). Finally, two attenuation plates and a BS are used to superposition these mixed states in a certain intensity proportion.
				
				$\rho_3$ can also be written as the mixture of two mixed  states
				\begin{equation}
					\begin{aligned}
						\rho_3 &= p \left |\psi_{\theta}\right \rangle \left\langle\psi_{\theta}\right| +(1-p)\rho_{\theta}^A\otimes I_2/2\\
						&=\frac{1+p}{2}\rho_{3a} + \frac{1-p}{2} \rho_{3b}.
					\end{aligned}\label{rho31}
				\end{equation}
			where  $\rho_{3a} = \cos^2 \theta |HH\rangle\langle HH| +\sin^2\theta|VV \rangle\langle  VV| +\frac{2p\cos \theta}{(1+p)\sin \theta}(|HH \rangle \langle VV| + |VV\rangle \langle HH|)$, $\rho_{3b} = \cos^2 \theta |HV  \rangle\langle  HV| +\sin^2 \theta|VH \rangle\langle VH | $, $p=0.55$ and $\theta = 0.3$. In Fig. \ref{setup}(\textbf{d}), we use two PPBSs to generate the state $\rho_{31} = (\cos^2 \theta|HH\rangle+ \sin^2\theta |VV\rangle)$.	Then we use BS split beam, the photon 2 in the upper path  passes through a quartz plate to partial decoherence and the photon 2 in the lower path passes through an HWP rotated at $45^{\circ}$ and a thick quartz plate to complete decoherence. Finally using two attenuation plates and a BS to superposition two mixed states in a ratio of $\frac{1+p}{1-p}$.

	 And $\rho_4$ can also be written as the mixture of two mixed  states
				\begin{equation}
					\begin{aligned}
						\rho_4&= \frac{1}{8}(\left | HH  \right \rangle \left \langle HH \right | +\left | VV  \right \rangle \left \langle VV \right| +2\left | HV  \right \rangle \left\langle HV \right|+4\left | \psi_+  \right \rangle \left\langle\psi_+\right|)\\
						&=	\frac{1}{4} \rho_{2a} +\frac{3}{4} \rho_{4a}.
					\end{aligned}\label{rho3}
				\end{equation}
				It can be generated by mixing the states $\rho_{2a}$ and $\rho_{4a} = \frac{2}{3}|HV  \rangle\langle HV| +\frac{1}{3}|HV \rangle\langle VH |+ \frac{1}{3}| VH \rangle\langle HV| +\frac{1}{3}| VH\rangle\langle VH| $ with a 1:3 light intensity ratio(which is shown in Fig. \ref{setup}(\textbf{e})). And $\rho_{4a} $ is generated by using a PPBS,  a HWP rotated at $45^{\circ}$ and a quartz plate to partial decoherence.

				Fig. \ref{setup}(\textbf{f}) shows the preparation of $\rho_5$
				\begin{equation}
					\rho_5 = (\left | HH \right \rangle \left\langle  HH \right| +\left | +V \right \rangle \left\langle +V \right| ) /2
				\end{equation}
				where $\ket{+} = (\ket{H}+\ket{V})/sqrt{2}$. The product state is prepared by selecting photon pairs produced from one BBO crystal by placing a PBS in the down path, while the classical mixture is achieved by inserting HWP in each path and randomly preparing the two product states with equal probability.

				Fig. \ref{setup}(\textbf{g}) shows the preparation of $\rho_6$, which is the same as $\rho_{2a}$.

				Fig. \ref{setup}(\textbf{h}) shows the preparation of $\rho_7$
				\begin{equation}
					\begin{aligned}
					\rho_7 &=(  \left | HH \right \rangle \left\langle HH \right| +  \left | VV  \right \rangle \left \langle VV \right| +  \left | + + \right \rangle \left\langle ++ \right| 	)/3\\
					&= \frac{2}{3} \rho_{2a} + \frac{1}{3}	\left | + + \right \rangle \left\langle ++ \right| .
					\end{aligned}
				\end{equation}	\label{rho7}
				The three HWPs angles from top to bottom are all rotated at $22.5^{\circ}$. The maximal mixing state($\rho_{2a}$) is generated by a HWP and a thick quartz plate in the BS's reflected light path. The state $\left | + + \right \rangle$ is generated by a PBS and a $22.5^{\circ}$ HWP in the BS's transmitted path. These two state is combined with a 2:1 light intensity ratio.

				Fig. \ref{setup}(\textbf{i}) shows the preparation of $\rho_8$
				\begin{equation}
					\begin{aligned}
						\rho_8 &= \frac{1}{4}(I_4+\frac{1}{3}\sigma _z\otimes I_2+\frac{1}{3}\sigma _x\otimes \sigma _x +\frac{1}{3}\sigma _y\otimes \sigma _y)\\
						&= \frac{1}{2}\rho_{8a}+\frac{1}{2}\rho_{8b}	
					\end{aligned}\label{rho8}
				\end{equation}
				Using a PPBS to let the transmittance of horizontal and vertical polarization photons is 2:1. And then respectively generated $\rho_{8a} = \frac{2}{3}|HH\rangle \langle HH| +\frac{1}{3}|VV \rangle \langle VV|$ and $\rho_{8b} = \frac{2}{3}|HV  \rangle\langle HV| +\frac{1}{3}|VH \rangle \langle VH|+ \frac{1}{3}| VH \rangle\langle  HV| +\frac{1}{3}| HV \rangle \langle VH| $ and combine them with same light intensity ratio.

				\subsection{Data analysis}
				
				To characterize the performance of our state preparations process, we do full-state tomography of two-qubit by using the same device in Fig.~\ref{setup}. The table~\ref{fidelity} shows the detailed fidelities of two qubits we have tested. Fig.~\ref{tom} shows the tomographic results for each state.

				\begin{figure*}[htbp]
					\centering
					\includegraphics[scale=0.045]{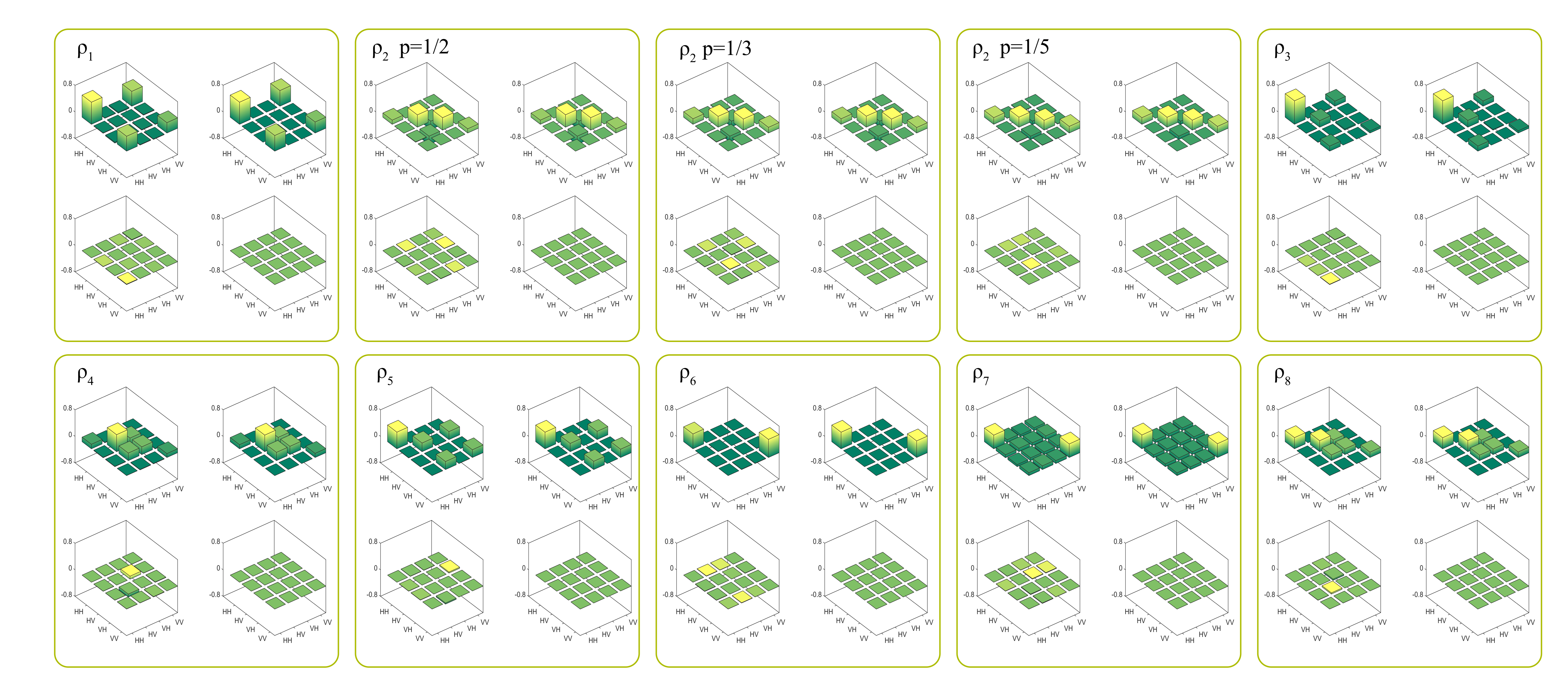}
					\caption{The states tomography results of two-qubit states($\rho_1$-$\rho_8$). In each box, the left two pictures show the real(top) and imaginary(bottom) parts of the experimental reconstructed density matrix, while the right two pictures show the theoretical density matrix. }\label{tom}
				\end{figure*}

				\begin{table}[htbp]
					\caption{\label{fidelity} The fidelities of all the two-qubits states we have tested, labeled by $\rho_1-\rho_8$. The error bars are determined by Monte Carlo simulation (50 samples) with the photonic statistic error.  }
					\begin{ruledtabular}
						\begin{tabular}{cccc}
							State&Fidelity&state&Fidelity\\
							\hline
							$\rho_1$	&$0.97084 \pm	 0.00058$	&$\rho_{21}$		&$0.99939  \pm   0.00005$	\\
							$\rho_{22}$	&$0.99916	\pm	0.00005$	&$\rho_{23}$		&$0.99946	\pm	0.00004$\\
							$\rho_3$	&$0.99779	\pm 0.00012$	& $\rho_4$ 			&$0.98723	\pm	 0.00014$\\
							$\rho_5$	&$0.99151	\pm 0.00046$	&$\rho_6$			&$0.99076	\pm	 0.00052$\\
							$\rho_7$	&$0.99320	\pm	 0.00070$	&$\rho_8$			&$0.99692	\pm 0.00010	$
						\end{tabular}
					\end{ruledtabular}
				\end{table}

				 Table~\ref{tab_wer} shows the center and semiaxes length of ellipsoids for each Werner state ($p=\frac{1}{2},\frac{1}{3}$ and $\frac{1}{5}$) and we can observe that for each $p$, the steering ellipsoid is nearly centered at the origin with three semiaxes approximately close to p.

				\begin{table}
				\caption{\label{tab_wer} The measured ellipsoid center and semiaxes length for each Werner state. The error bars are determined by Monte Carlo simulation with photonic statistics. }
				\begin{ruledtabular}
						\begin{tabular}{ccccccc}
								p& x0 &  y0 & z0&s1&s2&s3\\
								\hline
								$\mathcal{E}_{A|B}$	1/2&0.0120(4)	&	-0.0105(4)	&	0.0054(4)	&	0.5101(7)	&	0.5022(7)	&	0.4914(6)\\
								$\mathcal{E}_{B|A}$	1/2&0.0031(4)	&	-0.0039(4)	&	0.0011(4)	&	0.5090(5)	&	0.5015(5)	&	0.4906(6) \\
								$\mathcal{E}_{A|B}$	1/3&0.0135(4)  &	-0.0143(4) &  	0.0016(5)	&	0.3465(5)    &	0.3422(5)   & 	0.3218(5)\\
								$\mathcal{E}_{B|A}$	1/3&0.0045(4)  & 	-0.0064(3)  & 	-0.0025(4)	&	0.3473(6)    &	0.3394(5)    &	0.3196(6)\\
								$\mathcal{E}_{A|B}$	1/5&0.0002(4)  & 	-0.0047(4)   & 	0.0097(4)	&	 0.2125(7)   & 	0.2098(6)    &	0.1968(5)\\
								$\mathcal{E}_{B|A}$	1/5&0.0072(4)  &	-0.0112(4)   &	-0.0048(4)	&	0.2134(5)    &	0.2077(6)    &	0.1958(7)
							\end{tabular}
					\end{ruledtabular}
			\end{table}
		
	Table~\ref{tab_VAL} shows the results of fitted ellipsoid for states. We calculate the geometric properties of QSEs in $50$ experiments, i.e., the volume for $\rho_4$, area for $\rho_8$, and length for $\rho_6$. And we find that the ressults fit the predicted ellipsoids with high precision.	
		
					\begin{table}
					\caption{\label{tab_VAL} The geometric properties of fitted ellipsoid for states(which is ploted in Fig.~\ref{re_little}), i.e., the volume for $\rho_4$, area for $\rho_8$, and length for $\rho_6$. The error bars are determined by experiment data. exp. experimental; th, theoretical. }
					\begin{ruledtabular}
						\begin{tabular}{ccccc}
							State&$\mathcal{E}_{A|B}^{\rm exp}$&$\mathcal{E}_{A|B}^{\rm exp}$&$\mathcal{E}_{A|B}^{\rm th}$& $\mathcal{E}_{A|B}^{\rm th} $\\
							\hline
							$\rho_4$&$0.5133\pm 0.0077$	&$0.5539 \pm 0.0060$	&	0.5214	&	0.5214	\\
							$\rho_8$&$0.3573 \pm 0.0143$	&$0.4535 \pm 0.0920$	&	0.3490	&	0.3927\\
							$\rho_6$& $1.99995\pm 0.00004$ &$1.99999\pm  0.00001$ &  2	&2
						\end{tabular}
					\end{ruledtabular}
				\end{table}

				\clearpage
				
			\end{widetext}	
\end{document}